\documentclass[aps,pre,superscriptaddress,twocolumn,longbibliography]{revtex4}
\usepackage{graphicx,epstopdf,url}
\usepackage[colorlinks=true,urlcolor=blue,citecolor=blue,linkcolor=blue,urlcolor=blue]{hyperref}
\usepackage[active]{srcltx}
\usepackage{multirow}
\begin{document}
\title{
Thermal Stability of Two-Dimensional Crystals\\ with Extended OH Hydrogen-Bonded Chains
}
\author{Alexander V. Savin}
\email{asavin@chph.ras.ru}
\affiliation{
N.N. Semenov Federal Research Center for Chemical Physics of the Russian Academy of Sciences,
4 Kosygin St., Moscow 119991, Russia}
\affiliation{
Plekhanov Russian University of Economics, 36 Stremyanny Lane, Moscow 117997, Russia
}

\begin{abstract}
Numerical simulations of the dynamics of monolayer structures of molecules deposited on a sheet of hexagonal boron nitride (h-BN) have been performed.
It is shown that molecules containing benzene rings and hydroxyl groups can form stable two-dimensional crystals with linear hydrogen-bonded chains OH$\cdots$OH$\cdots$OH$\cdots$.
Such structures are formed by the following molecules: phenol (C$_6$H$_5$OH), hydroquinone (C$_6$H$_4$(OH)$_2$),
4-phenylphenol (C$_6$H$_5$--C$_6$H$_4$OH), 4-(4-phenylphenyl)phenol (C$_6$H$_5$--C$_6$H$_4$--C$_6$H$_4$OH),
paracetamol (CH$_3$C(O)NHC$_6$H$_4$OH), 4-hydroxybenzanilide (C$_6$H$_5$C(O)NHC$_6$H$_4$OH)
and 4,4$^\prime$-dihydroxybenzanilide (C$_6$H$_4$OHC(O)NHC$_6$H$_4$OH).
On the one hand, the benzene rings in these molecules ensure their strong interaction with the flat substrate; on the other hand, they do not hinder the formation of long hydrogen-bonded chains.
The monolayer structures of these molecules exhibit high thermal stability: the onset melting temperatures of their 2D crystals are 112, 257, 237, 337, 277, 407, and 483$^\circ$C, respectively.
The simulations suggest that structures composed of h-BN sheets and molecules of hydroquinone, paracetamol, 4-hydroxybenzanilide and 4,4$^\prime$-dihydroxybenzanilide can be used for the development of novel proton-exchange membranes capable of operating at elevated temperatures.

\end{abstract}

\maketitle
\section{Introduction}
The presence in a molecular system of chains of hydroxyl groups linked by hydrogen bonds,
\begin{equation}
\text{O--H}\cdots\text{O--H}\cdots\text{O--H}\cdots\text{O--H}\cdots\text{O--H}\cdots
\label{f1}
\end{equation}
ensures high proton conductivity in the direction of these chains \cite{Zundel2000}.
The transport of protons through cell membranes is carried out via protein proton channels and proceeds along hydrogen-bonded chains (\ref{f1}) formed by amino acid residues containing a hydroxyl group OH (serine, threonine, tyrosine) \cite{Nagle1978,Kaliman2008,Paulino2020}.
In the bacteriorhodopsin molecule, the hydrogen-bonded chain is formed by tyrosine residues contained in the seven transmembrane $\alpha$-helical segments of the molecule \cite{Merz1981}.
Hydrogen-bonded chains (\ref{f1}) act as proton conductors, providing an efficient pathway for fast proton transfer \cite{Fillaux2002}.
The idea of proton transfer along hydrogen-bonded chains was proposed by Theodor von Grotthuss as early as 1804 \cite{Grotthuss1806,Marx2006,Cukierman2006}.
According to modern concepts, proton transfer in water and ice occurs along a hydrogen-bond network and is divided into two phases: the passage of an ionic defect (H$^+$) and the passage of an orientational defect (Bjerrum defect), which, after the passage of the ionic defect, returns the hydrogen-bond chain to its original state \cite{Bjerrum1952,Nagle1978,Merz1981}.
The highest proton conductivity is possessed by phosphoric acid H$_3$PO$_4$, which is capable of forming branched hydrogen-bonded chains \cite{Viliauskas2012}.

At present, the creation of molecular systems with high proton conductivity is an urgent task for the development and improvement of proton-exchange membranes (PEMs), whose main purpose is to conduct protons (H$^+$) while simultaneously serving as an insulator for electrons and a barrier for gases \cite{Kiani2025,Luo2026}.
PEM fuel cells have demonstrated high potential as environmentally clean alternative energy sources.
However, their use is constrained by the degradation of PEM performance at high temperatures and low humidity.
To solve this problem, it is necessary to create new materials with high proton conductivity at elevated temperatures and in the absence of water.
Such materials must consist of molecules capable of creating hydrogen-bonded chains (\ref{f1}) that are stable at high temperatures.

In this work, using molecular dynamics, we will analyze the possibility of creating such materials from flat molecules deposited on a sheet of hexagonal boron nitride (h-BN).
These molecules must contain benzene rings and hydroxyl groups OH in their structure
(the rings ensure strong interaction of the molecules with the flat substrate, while the hydroxyl groups provide the possibility of forming extended hydrogen-bonded chains necessary for proton transport).
Structures made of phenol, hydroquinone, paracetamol, and their more complex analogues will be considered.

\section{Model}

\begin{figure}[tb]
\begin{center}
\includegraphics[angle=0, width=0.8\linewidth]{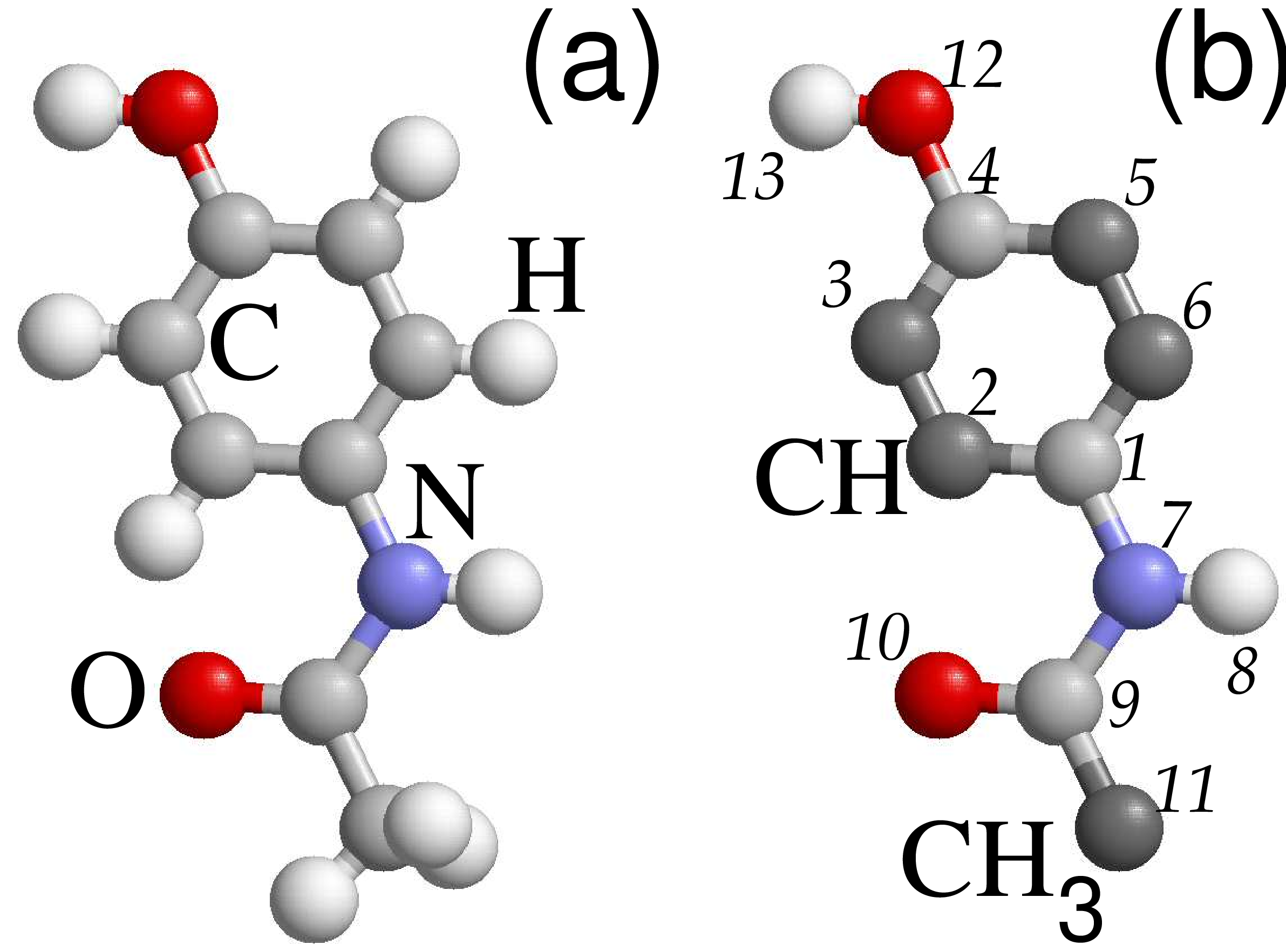}
\end{center}
\caption{\label{fig01}\protect
Construction of the coarse-grained model for the paracetamol (PCM) molecule: (a) all-atom and (b) coarse-grained model (the atom numbering is shown).
Hydrogen atoms are shown in white, carbon in light gray, nitrogen in blue, oxygen in red, and the united CH and CH$_3$ atoms in dark gray.
}
\end{figure}

In the simulations, we use the united-atom approximation, in which the CH and CH$_3$ groups are treated as united atoms centered on the carbon atoms.
As an example, let us illustrate the construction of the coarse-grained model for the paracetamol molecule CH$_3$C(O)NHC$_6$H$_4$OH (PCM) (Fig. \ref{fig01}(a)).
In this approximation, the PCM molecule is represented as a system of $N_0=13$ united atoms (Fig. \ref{fig01}(b)).
The masses of the united atoms are listed in Table \ref{tab1}.

To model the PCM molecule, we use a force field in which various potentials describe bond deformations of valence bonds, valence and dihedral (torsion) angles, as well as nonbonded interactions of atoms.
In this model, the deformation energy of the valence bonds C--CH,\linebreak CH--CH, C--N, N--H, C=O, C--CH$_3$, C--O, and O--H is described by a harmonic potential
\begin{equation}
U_{\text{v}}(\rho)=\frac12K(\rho-\rho_0)^2,
\label{f2}
\end{equation}
where $\rho$ and $\rho_0$ are the current and equilibrium bond lengths, and $K$ is the bond stiffness.
The parameters of potential (\ref{f2}) for various valence bonds are given in Table \ref{tab2}.
\begin{table*}[tb]
\caption{
Masses and parameters of the interaction potentials for the united atoms X of the PCM molecule: $i$ --- atom number; $M_i$ --- atom mass ($m_p=1.6603\times 10^{-27}$~kg --- proton mass);
$\varepsilon_i$ and $r_i$ --- energy and radius of the LJ interaction; $q_i$ --- electric charge of the atom; $\epsilon_i$ and $h_i$ --- interaction energy and equilibrium distance from the flat substrate (surface of the h-BN crystal).
\label{tab1}
}
\begin{center}
\setlength{\tabcolsep}{7pt}
\begin{tabular*}{0.83\linewidth}{ccccccccccc}
\hline\hline
 X                 &   C   & C   & CH       &  N   & H   & C   & O    & CH$_3$ & O    &  H\\
 $i$               &   1   & 4   &2, 3, 5, 6&  7   & 8   & 9   & 10   & 11     & 12   & 13\\
\hline
 $M_i$~($m_p$)     &  12   & 12  & 13       & 14   & 1   & 12  & 16   & 15     & 16   &  1\\
$\varepsilon_i$~(meV)&4.284&4.284&4.284     &4.080 &0.434&4.284&6.344 &4.284   &6.344 &0.434\\
$r_i$~(\AA)        &1.861  &1.861&1.861     &1.899 &0.621&1.861&1.711 &1.861   &1.711 &0.621\\
$q_i$~(e)          &0.066  &0.100& 0        &-0.463&0.286&0.580&-0.504&0.035   &-0.500&0.400\\
$\epsilon_i$~(meV) &82.2   &82.2 &120.0     &101.1 &29.1 & 82.2& 91.2 & 157.8  & 77.8 &29.1\\
$h_i$~(\AA)        &3.26   &3.26 &3.26      &3.20  &2.91 & 3.26& 3.13 & 3.64   & 3.23 &2.91\\
\hline\hline
\end{tabular*}
\end{center}
\end{table*}
\begin{table}[tb]
\caption{
Values of the parameters of the harmonic potential (\ref{f2}) for various valence bonds X--Y
\label{tab2}
}
\begin{center}
\setlength{\tabcolsep}{2.5pt}
\begin{tabular}{cccccccc}
\hline\hline
 X--Y          & C--CH & C--N &  N--H  & C=O  & C--CH$_3$ & C--O & O--H \\
  ~            & CH--CH& ~    &   ~    &  ~  & ~ & ~ & ~ \\
\hline
 $K$~(N/m)     & 469           & 427  & 434    & 570  & 553      & 450  & 317\\
 $\rho_0$~(\AA)& 1.39          &1.405 & 1.007  &1.222 & 1.505    & 1.364& 0.96\\
\hline\hline
\end{tabular}
\end{center}
\end{table}

The valence-angle bending energy for angles X--Y--Z is described by the potential
\begin{equation}
U_{\text{a}}({\mathbf u}_1,{\mathbf u}_2,{\mathbf u}_3)=\varepsilon_a(\cos\varphi-\cos\varphi_0)^2,
\label{f3}
\end{equation}
where $\cos\varphi=-{\mathbf v}_1\cdot{\mathbf v}_2/\rho_1\rho_2$ is the cosine of the valence angle,
the vectors ${\mathbf v}_1={\mathbf u}_2-{\mathbf u}_1$, ${\mathbf v}_2={\mathbf u}_3-{\mathbf u}_2$,
and the valence bond lengths $\rho_1=|{\mathbf v}_1|$, $\rho_2=|{\mathbf v}_2|$.
Here the vectors ${\mathbf u}_1$, ${\mathbf u}_2$, and ${\mathbf u}_3$ specify the coordinates of the atoms forming the valence angle $\varphi$, and $\varphi_0$ is the equilibrium value of the angle.
The potential parameters for the various valence angles are listed in Table~\ref{tab3}.

The torsional (dihedral) energy is described by the potential
\begin{equation}
U_{\text{d}}({\mathbf u}_1,{\mathbf u}_2,{\mathbf u}_3,{\mathbf u}_4)=\epsilon_d(1+z_d\cos\phi),
\label{f4}
\end{equation}
where $\cos\phi={\mathbf w}_1\cdot{\mathbf w}_2/|{\mathbf w}_1||{\mathbf w}_2|$,
the vectors ${\mathbf w}_1=({\mathbf u}_2-{\mathbf u}_1)\times ({\mathbf u}_3-{\mathbf u}_2)$ and
${\mathbf w}_2=({\mathbf u}_3-{\mathbf u}_2)\times ({\mathbf u}_4-{\mathbf u}_3)$.
The parameters used for various dihedral angles are listed in Table \ref{tab4}.

For a pair of atoms X$_i$, X$_j$ ($i$, $j$ are the atom numbers in the molecule) participating in the formation of the dihedral angle X$_i$--Y--Z--X$_j$, their nonbonded interaction is also taken into account, described by the Lennard-Jones (LJ) potential
\begin{equation}
U_{\mathrm{LJ}}(r)=\varepsilon_0[(r_0/r)^{12}-2(r_0/r)^6],
\label{f5}
\end{equation}
with the interaction energy $\varepsilon_0=\sqrt{\varepsilon_i\varepsilon_j}/2$, where
$r$ is the current distance between the interacting atoms, and the equilibrium distance $r_0=r_i+r_j$.
The LJ interaction of the oxygen atom of the amide group ($i=10$) with the united CH atoms ($i=2$, 6) is also taken into account, with the interaction energy $\varepsilon_0=\sqrt{\varepsilon_2\varepsilon_{10}}$ and the equilibrium distance $r_0=r_2+r_{10}$.
The parameters $\varepsilon_i$ and $r_i$ are listed in Table~\ref{tab1}.

The interaction of two PCM molecules is described by the potential
\begin{equation}
\setlength{\medmuskip}{0.5mu}
\setlength{\thickmuskip}{0.5mu}
U({\mathbf X}_1,{\mathbf X}_2)=\sum_{i=1}^{N_0}\sum_{j=1}^{N_0}\left\{\varepsilon_{ij}
\left[\left(\frac{\bar{r}_{ij}}{r_{ij}}\right)^{12}-2\left(\frac{\bar{r}_{ij}}{r_{ij}}\right)^{6}\right]+\kappa \frac{q_iq_j}{r_{ij}}\right\},
\label{f6}
\end{equation}
where $N_0=13$ is the number of united atoms in the molecule, and the $3N_0$-dimensional vector
${\mathbf X}_k=\{{\mathbf u}_{k,i}\}_{i=1}^{N_0}$ ($k=1,2$) specifies the coordinates of the atoms of the molecule (the vector ${\mathbf u}_{k,i}$ specifies the position of the $i$-th atom of the $k$-th molecule), and the distance between atoms is $r_{ij}=|{\mathbf u}_{1,i}-{\mathbf u}_{2,j}|$.
Here, the energy $\varepsilon_{ij}=\sqrt{\varepsilon_i\varepsilon_j}$, the equilibrium distance
$\bar{r}_{ij}=r_i+r_j$, and $q_i$ is the electric charge of the atom ($i,j=1,\dots,N_0$), with the coefficient
$\kappa=14.400611$~eV$\cdot$\AA/e$^2$.
The parameters $\varepsilon_i$, $r_i$, and $q_i$ are listed in Table \ref{tab1}.
All parameter values of the interaction potentials (\ref{f2})--(\ref{f6}) were obtained 
from the General AMBER Force Field (version 2.1, April 2016) \cite{Amber}.

In the simulations, we use the approximation of a fixed attractive plane for the substrate.
In this approximation, the van der Waals interaction of the atoms of the molecule with the flat substrate can be described by the LJ potential $(m,l)$
\begin{equation}
\setlength{\medmuskip}{0.5mu}
\setlength{\thickmuskip}{0.5mu}
W({\mathbf X})=\sum_{i=1}^{N_0}W_i(z_i)=\sum_{i=1}^{N_0}\frac{\epsilon_i}{l-m}
\left[ m\left(\frac{h_i}{z_i}\right)^l-l\left(\frac{h_i}{z_i}\right)^m\right],
\label{f7}
\end{equation}
where $z_i$ is the distance from the $i$-th atom to the outer surface of the flat substrate $z\le 0$.
The potential $W_i(z_i)$ in Eq. (\ref{f7}) gives the dependence of the interaction energy of the $i$-th atom on its distance from the substrate. 
This dependence was obtained numerically for various substrates \cite{Savin2019,Savin2021}.
The potential $W_i(z_i)$ has a minimum $W_i(h_i)=-\epsilon_i$ ($\epsilon_i$ is the binding energy of the atom to the substrate).
As the flat substrate, we will use the surface of an h-BN crystal, for which
the exponents are $l=10$, $m=3.75$.
The parameters $\{\epsilon_i, h_i\}_{i=1}^{N_0}$ are listed in Table \ref{tab1}.
Note that an h-BN sheet, unlike a graphene sheet, is not a conductor.
Therefore, when constructing structures for proton-exchange membranes, h-BN is more preferable.

\begin{figure}[tb]
\begin{center}
\includegraphics[angle=0, width=1.\linewidth]{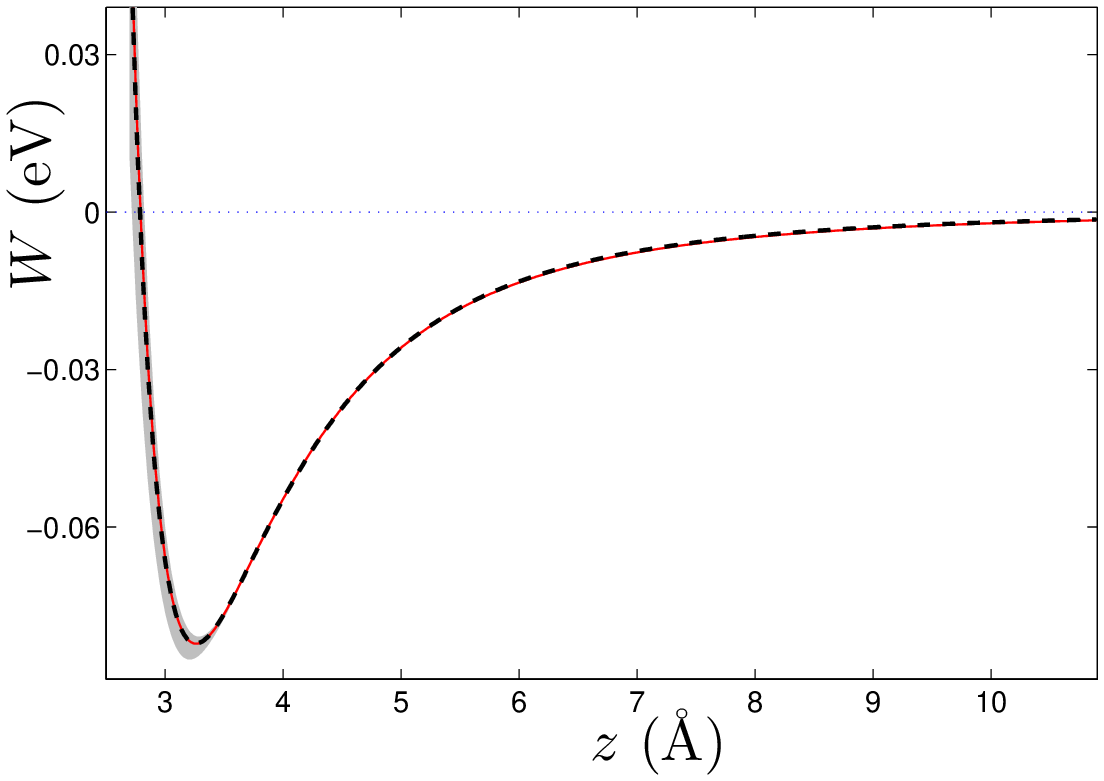}
\end{center}
\caption{\label{fig02}\protect
Interaction energy $W(z)$ of a carbon atom with the flat surface ($z=0$) of an h-BN crystal as a function of the distance $z$ from the surface.
 The solid red curve shows the numerically obtained dependence averaged over the unit cell of the top sheet of the crystal. 
The dashed line shows the LJ potential (\ref{f9})   with $\epsilon_0=0.0822$~eV and $h_0=3.26$~\AA.
The gray band around the curve shows the range of variation of the unaveraged dependence.
}
\end{figure}

The interaction of a particle with the half-space $z\le 0$ is usually described by the LJ potential (3,9)
\begin{equation}
W(z)=\epsilon_0[(h_0/z)^9-3(h_0/z)^3]/2, \label{f8}
\end{equation}
where $\epsilon_0$ is the interaction energy (adhesion energy), and $h_0$ is the equilibrium distance from the half-space surface \cite{Aitken2010,Zhang2013,Zhang2014}.
The interaction of a particle with the surface of an h-BN crystal is more accurately described by the LJ potential
\begin{equation}
W(z)=\epsilon_0(l-m)[m(h_0/z)^l-l(h_0/z)^m], \label{f9}
\end{equation}
with exponents $l=10$, $m=3.75$ --- see Fig. \ref{fig02}.
The difference between potentials (\ref{f8}) and (\ref{f9}) is due to the fact that the substrate has a layered structure and the main contribution comes from the interaction with the first layer (i.e., with the top sheet of the crystal).
\begin{table}[tb]
\caption{
Values of the parameters for the valence angle potential X--Y--Z (\ref{f3}) for different atoms
\label{tab3}
}
\begin{center}
\setlength{\tabcolsep}{1.5pt}
\begin{tabular}{cccccccccc}
\hline\hline
 XYZ          & CCC & CCN & CNH & CNC & NCO & NCC & OCC & CCO & COH\\
\hline
$\varepsilon_a$~(eV)   & 3.643&3.823    & 2.781  & 4.888 &4.932   & 3.758  & 4.625  & 4.047  & 1.791\\
$\varphi_0$~($^\circ$)&120 & 117     &  118   & 128   & 123    & 116    & 120    & 120    & 113\\
\hline\hline
\end{tabular}
\end{center}
\end{table}

The potential (\ref{f9}) was obtained numerically as the dependence of the interaction energy of a particle
(a united atom) with a 10-layer h-BN sheet on the distance $z$ from the sheet surface.
The calculations employed sheets of size $5.06\times 4.89$~nm$^2$ consisting of 960 atoms.
The surface of the top sheet coincided with the plane $z=0$, and the particle was located above the center of the  sheet at a distance $z>0$.
The interaction energy $W(z)$ was calculated as the sum of all pairwise interactions between the particle and the atoms of the multilayer sheet using the LJ potentials (\ref{f5}) with parameter values from the AMBER force field \cite{Amber} (the parameter set from version GAFF2 was used).
The dependence $W(z)$ was averaged over the unit cell of the sheet in the $x$ and $y$ directions.
As can be seen from Fig. \ref{fig02}, the numerically obtained dependence $W(z)$ is well described  by the potential (\ref{f9}).
\begin{table*}[tb]
\caption{
Values of the parameters for the dihedral angle potential X--Y--Z--W (\ref{f4}) for different atoms
\label{tab4}
}
\begin{center}
\setlength{\tabcolsep}{5pt}
\begin{tabular*}{0.82\linewidth}{cccccccccc}
\hline\hline
 XYZW            & CCCC & CCCN & C6C1NH & C2C1NH & C6C1NC &C2C1NC & CNCO & CNCC11 & CCCO\\
\hline
$\epsilon_d$~(eV)& 0.63 & 0.63 & 0.42   & 0.42   & 0.42 & 0.42    & 0.42  & 0.42  & 0.63\\
$z_d$            & -1   & 1    &  -1    &  1     &  1   & -1      & -1    &   1   & 1\\
\hline\hline
\end{tabular*}
\end{center}
\end{table*}

Note that the interaction energy of a particle depends only weakly on its coordinates $x$, $y$.
The variations of the interaction energy across the sheet do not exceed $5$\%.
In Fig. \ref{fig02} this range of variation is shown as a narrow gray band around the averaged curve $W(z)$).
Therefore, the flat molecules we simulated here perceive the substrate as an almost homogeneous attractive plane, the interaction with which is described by the sum of potentials $W_i(z_i)$, (\ref{f7}).
The dynamics of the system of molecules is then simulated using the Langevin equations, in which random forces model the interaction of the molecules with thermal vibrations of the substrate. 
A fully atomistic substrate model would not reveal any new effects but would lead to a significant increase in the dimensionality of the numerically integrated system.
\begin{figure}[tb]
\begin{center}
\includegraphics[angle=0, width=0.8\linewidth]{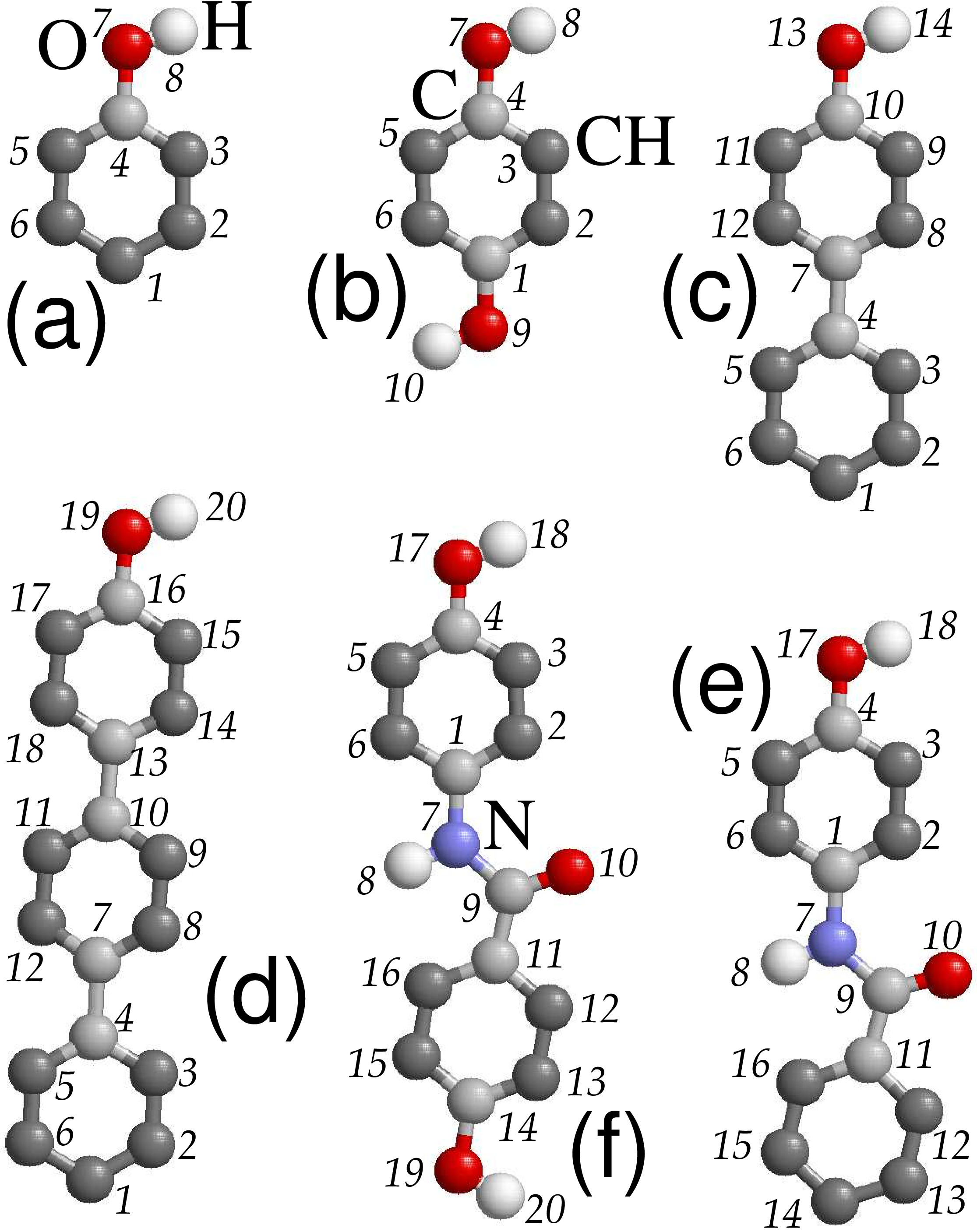}
\end{center}
\caption{\label{fig03}\protect
Coarse-grained models of (a) phenol molecule C$_6$H$_5$OH,
(b) hydroquinone C$_6$H$_4$(OH)$_2$,
(c) 4-phenylphenol C$_6$H$_5$--C$_6$H$_4$OH,
(d) 4-(4-phenylphenyl)phenol C$_6$H$_5$--C$_6$H$_4$--C$_6$H$_4$OH,
(e) 4-hydroxybenzanilide C$_6$H$_5$C(O)NHC$_6$H$_4$OH and
(f) 4,4$^\prime$-dihydroxybenzanilide C$_6$H$_4$OHC(O)NHC$_6$H$_4$OH.
Hydrogen atoms H are shown in white, carbon C in light gray, nitrogen N in blue,
oxygen O in red, and the united CH atoms in dark gray.
For each structure, the atom numbering is shown.
}
\end{figure}

The same united-atom model can also be constructed for other molecules containing benzene rings, hydroxyl OH groups, and amide (peptide) HNCO groups (see Fig. \ref{fig03}).
These molecules consist of the same united atoms and the same bonds as  paracetamol; therefore the same potentials and parameter sets can be used in modeling them.
Note only that the valence bond connecting the benzene rings in 4-phenylphenol
and 4-(4-phenylphenyl)phenol (see Fig.~\ref{fig03}(c) and (d)) is a single bond, and the planes of adjacent rings in an isolated molecule form a dihedral angle $\phi_t=134^\circ$.
Since this equilibrium angle differs from 0$^\circ$ and 180$^\circ$, potential (\ref{f4}) is insufficient; instead, to describe the deformation of the dihedral angles formed by the atoms C3--C4--C7--C12 and C5--C4--C7--C8, the potential
\begin{equation}
U_t({\mathbf u}_1,{\mathbf u}_2,{\mathbf u}_3,{\mathbf u}_4)=\epsilon_t(\cos\phi-\cos\phi_t)^2,
\label{f10}
\end{equation}
should be used, where $\epsilon_t=0.045$~eV is the energy parameter \cite{Johansson2008}.

Benzene rings ensure strong interaction of the molecules with the flat substrate, while hydroxyl and amide groups enable hydrogen bonding between molecules. 
Let each molecule consist of $N_0$ united atoms.
The Hamiltonian of a system of $N$ molecules located on a flat substrate has the form
\begin{equation}
{\mathcal H}=\sum_{n=1}^N\frac12{\mathbf M}\dot{\mathbf X}_n\cdot\dot{\mathbf X}_n+P,
\label{f11}
\end{equation}
where the first term is the kinetic and the second is the potential energy of the system,
\begin{equation}
P=\sum_{n=1}^N[V({\mathbf X}_n)+W({\mathbf X}_n)]+\sum_{n=1}^{N-1}\sum_{k=n+1}^N U({\mathbf X}_n,{\mathbf X}_k).
\label{f12}
\end{equation}
Here the vector ${\mathbf X}_n=\{ {\mathbf u}_{n,i}\}_{i=1}^{N_0}$ gives the coordinates of the atoms of the $n$-th molecule, and ${\mathbf M}$ is the diagonal mass matrix, $V({\mathbf X}_n)$ is the deformation energy of $n$-th molecule, $W({\mathbf X}_n)$ is its interaction energy with substrate, and $U({\mathbf X}_n,{\mathbf X}_k)$ is the interaction energy of molecules $n$ and $k$.

To find the stationary state of a system of $N$ molecules located on a flat h-BN substrate,
it is necessary to find the state of the system with the minimum potential energy
\begin{equation}
P\rightarrow\min:\{ {\mathbf X}_n\}_{n=1}^N,
\label{f13}
\end{equation}
where $P$ is defined by (\ref{f12}).
The minimization problem (\ref{f13}) was solved numerically using the conjugate gradient method \cite{Fletcher1964,Shanno1976}.
By varying the initial configurations of the molecules in the minimization procedure, one can obtain all ground-state configurations of the molecular system.

\section{Two-Dimensional Crystals of Paracetamol Molecules}

\begin{figure}[tb]
\begin{center}
\includegraphics[angle=0, width=1.0\linewidth]{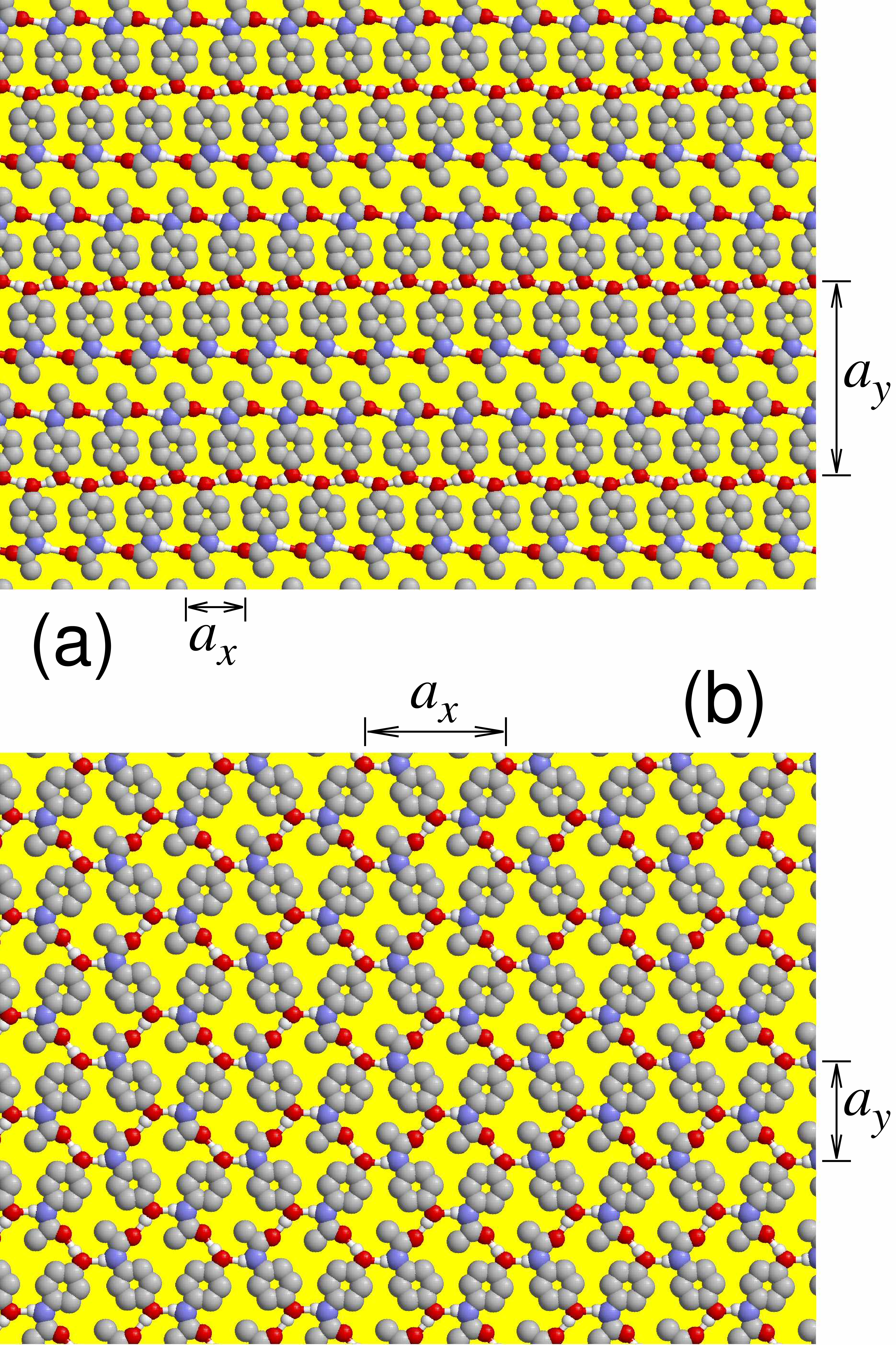}
\end{center}
\caption{\label{fig04}\protect
Crystal structure of (a) the homochiral and (b) racemic layer of paracetamol molecules on the flat surface of an h-BN crystal.
Carbon atoms are shown in gray, nitrogen in blue, oxygen in red, hydrogen in white;
the flat substrate is shown in yellow.
Hydrogen bonds between hydroxyl and amide groups are shown as red-white lines.
The lattice periods $a_x$ and $a_y$ are indicated.
}
\end{figure}

Numerical minimization of the potential energy, Eq.~(\ref{f13}), showed that PCM molecules on the flat h-BN substrate can form two types of two-dimensional periodic (crystalline) structures (see Fig.~\ref{fig04}).
In these structures, each molecule forms four hydrogen bonds (two incoming and two outgoing).
In the first structure, the hydroxyl (hg) and amide (ag) groups of the molecules form linear chains of hydrogen bonds (\ref{f1}) and
\begin{equation}
\text{H--N--C=O}\cdots\text{H--N--C=O}\cdots\text{H--N--C=O}\cdots
\label{f14}
\end{equation}
(Fig. \ref{fig04}(a)).
Thus, the periodic structure consists of parallel molecular chains that form three hydrogen-bonded chains: two chains of type (\ref{f14}) and one chain of type (\ref{f1}) located between them.

In the second structure, only zigzag chains of mixed hydrogen bonds are formed
\begin{equation}
\text{H--N--C=O}\cdots\text{H--O}\cdots\text{H--N--C=O}\cdots\text{H--O}\cdots
\label{f15}
\end{equation}
(Fig. \ref{fig04}(b)).
Here each molecule is linked by hydrogen bonds to all its neighbors.
This hydrogen-bond topology is characteristic of the three-dimensional PCM crystal \cite{Boldyreva2004,Anitha2015}.
\begin{figure}[tb]
\begin{center}
\includegraphics[angle=0, width=1.0\linewidth]{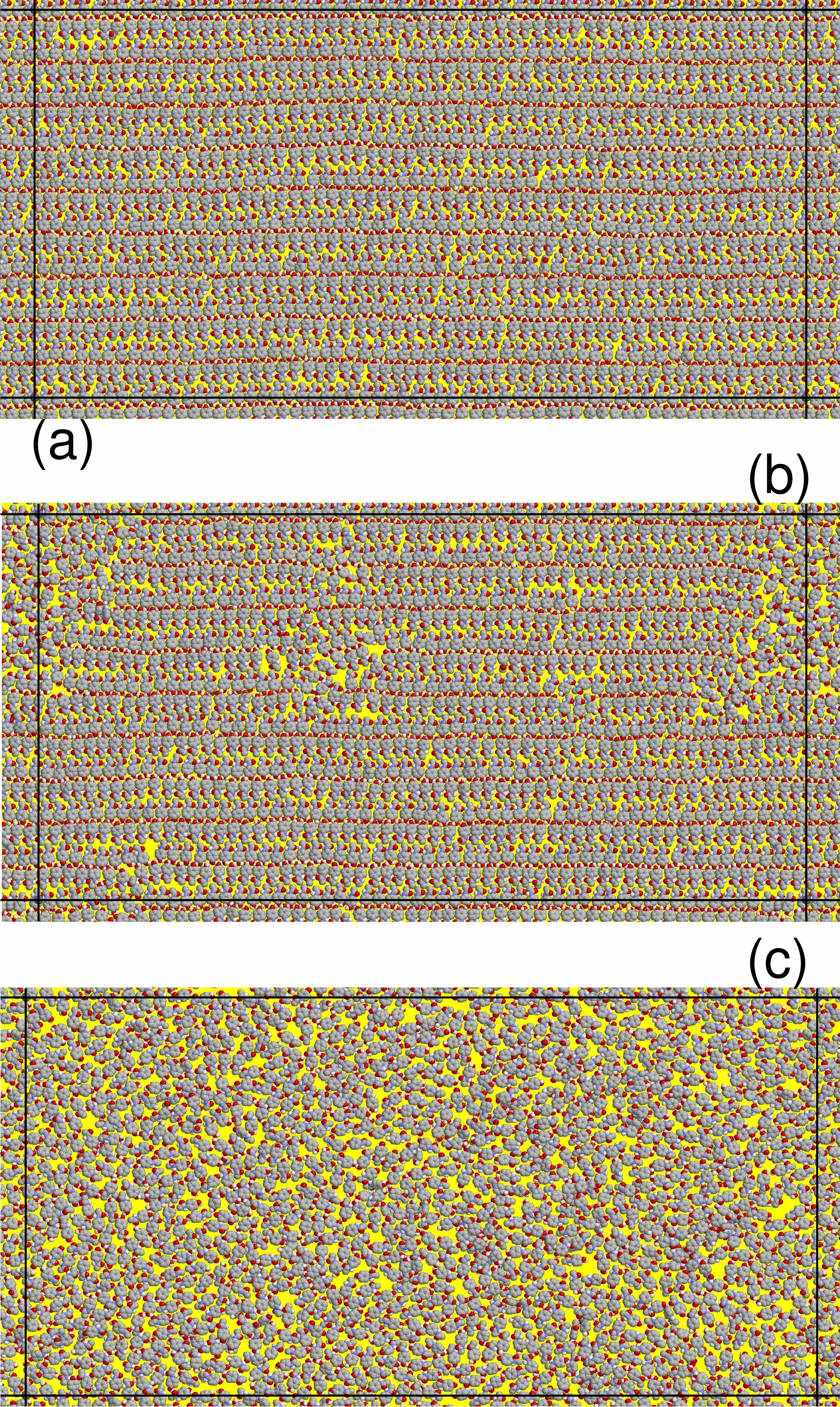}
\end{center}
\caption{\label{fig05}\protect
Structure of the 2D crystal of $N=1080$ PCM molecules at 
(a) $T=540$~K (onset of melting; $p_{\mathrm{hh}}=0.84$), 
(b) 560~K, and 
(c) 580~K (end of melting, $p_{\mathrm{hh}}=0.15$).
The periodic cell is outlined in black.
}
\end{figure}
\begin{figure}[tb]
\begin{center}
\includegraphics[angle=0, width=1.0\linewidth]{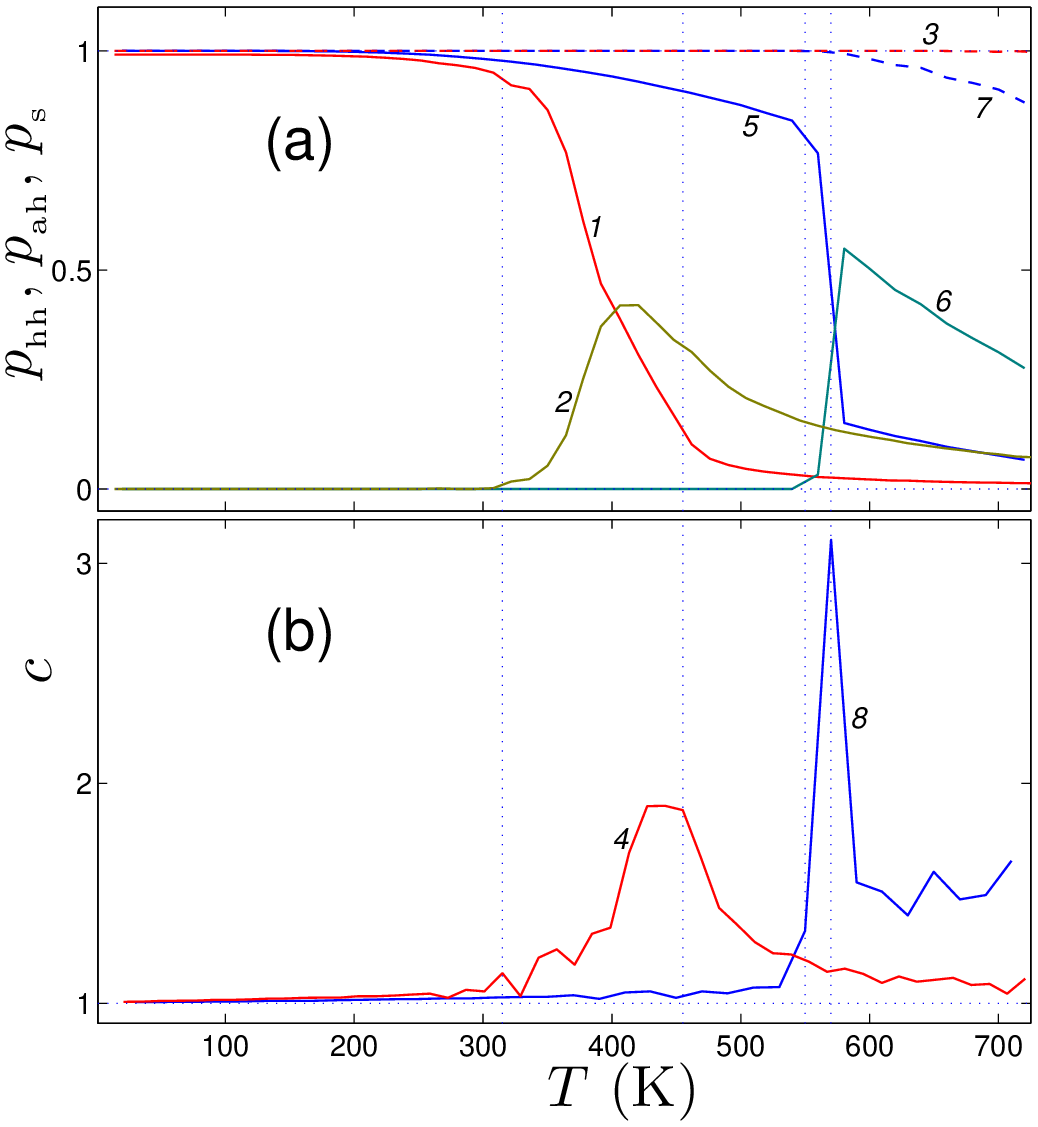}
\end{center}
\caption{\label{fig06}\protect
Temperature dependencies of (a) the normalized numbers of hydrogen bonds $p_{\mathrm{hh}}$ and $p_{\mathrm{ah}}$ (solid lines: 1, 5 for $p_{\mathrm{hh}}$  and 2, 6 for $p_{\mathrm{ah}}$) and the fraction $p_{\mathrm s}$ of molecules lying on the substrate (dashed lines 3 and 7), and (b) the dimensionless heat capacity $c$ (curves 4 and 8), for the homochiral structure of $N=1080$ PCM molecules.
Curves 1, 2, 3, and 4 correspond to a rectangular crystallite with free edges, covering 25~\% of the substrate, while curves 5, 6, 7, and 8 correspond to a two-dimensional crystal covering the entire substrate.
The vertical dotted lines mark the temperatures 315, 455, 550, and 570~K.
}
\end{figure}

The PCM molecule is achiral, but it becomes chiral when placed on a flat substrate.
Depending on which side it lies on the substrate, it can be either right-handed (when the benzene ring lies to the right of the vector $\overrightarrow{\mathrm{OH}~}$ connecting the terminal atoms of the amide group) or left-handed (otherwise) \cite{Savin2023}.
As seen in Fig.~\ref{fig04}, in the first crystalline structure all molecules have the same chirality, so this structure is homochiral.
In the second structure, half of the molecules are left-handed enantiomers and half are right-handed,
so this structure is racemic.
The unit cell of each 2D crystal contains two molecules.
The periods of the first lattice are $a_x=5.51$, $a_y=18.47$~\AA, and those of the second are $a_x=12.18$, $a_y=8.55$~\AA.
Both structures are stable and have similar packing densities: in the homochiral structure, the area per molecule is $S_0=50.88$~\AA$^2$, and in the racemic one, $S_0=52.07$~\AA$^2$.
The second structure is energetically more favorable, with an energy difference per molecule of 
$\Delta E=-0.1$~eV.
Note that this structure is not a proton conductor, since it lacks hydrogen-bonded chains of type (\ref{f1}).

To simulate the dynamics of the first (homochiral) structure, let us consider a 2D crystal of $N=2N_xN_y$ PCM molecules, where the factor 2 accounts for the two molecules per unit cell. 
The crystal consisting of $N_y=9$ parallel chains, each containing $N_x=60$ units (see Fig. \ref{fig05}).
When periodic boundary conditions with periods $L_x=N_xa_x$ and $L_y=N_ya_y$ are used, this structure models an infinite two-dimensional crystal densely covering the entire substrate.
When a periodic cell of size $L_x=2N_xa_x$, $L_y=2N_ya_y$ is used, a two-dimensional crystallite with free edges is modeled, covering only 25\% of the substrate surface.

Numerical minimization of the potential energy, Eq.~(\ref{f13}), showed that all molecular chains forming the 2D crystal are linear (see Fig.~\ref{fig05}(a)).
The hydrogen-bonded chains of hydroxyl groups (\ref{f1}) have the form of a zigzag chain with a distance
between neighboring oxygen atoms $|\mathrm{OO}|=2.88$~\AA~ and an angle $\angle\mathrm{OOO}=146^\circ$.
The hydrogen bond length is $|{\mathrm O}\cdots{\mathrm H}|=1.90$~\AA, and the bond energy is $E_{{\mathrm{hh}},0}=-0.23$~eV.
The hydrogen-bonded chains of amide groups (\ref{f14}) are linear with a step $a=a_x=5.51$~\AA.
The hydrogen bond length between amide groups is $|{\mathrm O}\cdots{\mathrm H}|=2.38$~\AA,
and the bond energy is $E_{{\mathrm{aa}},0}=-0.21$~eV.
The hydrogen bond energy between hydroxyl groups, $E_{\mathrm{hh}}$, can be defined as the interaction energy of atoms with numbers $i\in I_{\mathrm h}$ (in one molecule),
and the bond energy between amide groups $E_{\mathrm{aa}}$ as the interaction energy of atoms with numbers
\linebreak
$i\in I_{\mathrm a}$, where the sets of index values are $I_{\mathrm h}=\{3, 4, 5, 12, 13\}$,
$I_{\mathrm a}=\{1, 2, 6,...,11\}$ (for this end, in Eq.~(\ref{f6}) the summation should be performed only over these values of $i$ and $j$).

Let us place the molecular structure in a Langevin thermostat and consider its dynamics.
To this end, we numerically integrate the system of Langevin equations
\begin{equation}
{\mathbf M}\ddot{\mathbf X}_n=-\frac{\partial\mathcal H}{\partial {\mathbf X}_n}-\Gamma {\mathbf M}\dot{\mathbf X}_n +\Xi_n,~n=1,\dots,N,
\label{f16}
\end{equation}
with initial conditions corresponding to the stationary state of the molecular structure.
Here $\mathcal H$ is the Hamiltonian of the system, Eq.~(\ref{f11});
$\Gamma=1/t_r$ is the friction coefficient characterizing the intensity of energy exchange with the thermostat (relaxation time $t_r=10$~ps);
$\Xi_n=\{\xi_{n,i,k}\}_{i=1,k=1}^{N_0,~3}$ is a $3N_0$-dimensional vector of normally distributed random forces normalized by the conditions
\[
\langle\xi_{n,i,k}(t_1)\xi_{m,j,l}(t_2)\rangle=2M_nk_BT\Gamma\delta_{nm}\delta_{ij}\delta_{kl}\delta(t_2-t_1)
\]
($T$ is the thermostat temperature, $k_B$ is the Boltzmann constant).

In the simulated molecular system, the role of the thermostat is played by the flat substrate;
the random forces in the system of equations of motion (\ref{f16}) model the interaction of molecules with the thermal vibrations of the substrate.
The relaxation time $t_r$ determines the intensity of energy exchange with the thermostat.
The interaction with the substrate is governed by weak nonbonded forces.
For such interactions, the relaxation time (the energy-exchange time with the thermostat) is
$t_r=0.1$--1~ns.
For computational convenience we use a smaller value, $t_r=10$~ps.
This allows us to significantly reduce the integration time required to reach the equilibrium
of the molecular system with the thermostat and to obtain reliable average values.

The system of equations of motion (\ref{f16}) was integrated numerically by the velocity Verlet method \cite{Verlet1967} with a constant time step $\Delta t=1$~fs.
The initial conditions were the equilibrium configurations of the molecular system obtained by solving the energy-minimization problem, Eq.~(\ref{f13}). 
The calculations were performed on the supercomputer of the Joint Supercomputer Center (JSCC) of the National Research Center ``Kurchatov Institute''. 
Each node performed integration at a fixed thermostat temperature, so that thermalized states of the molecular system were obtained simultaneously for different temperatures. 
All programs were written by the author independently in Fortran.
Only for solving the minimization problem (\ref{f13}) the  CONMIN program \cite{Shanno1976} was used.
\begin{figure}[tb]
\begin{center}
\includegraphics[angle=0, width=1.0\linewidth]{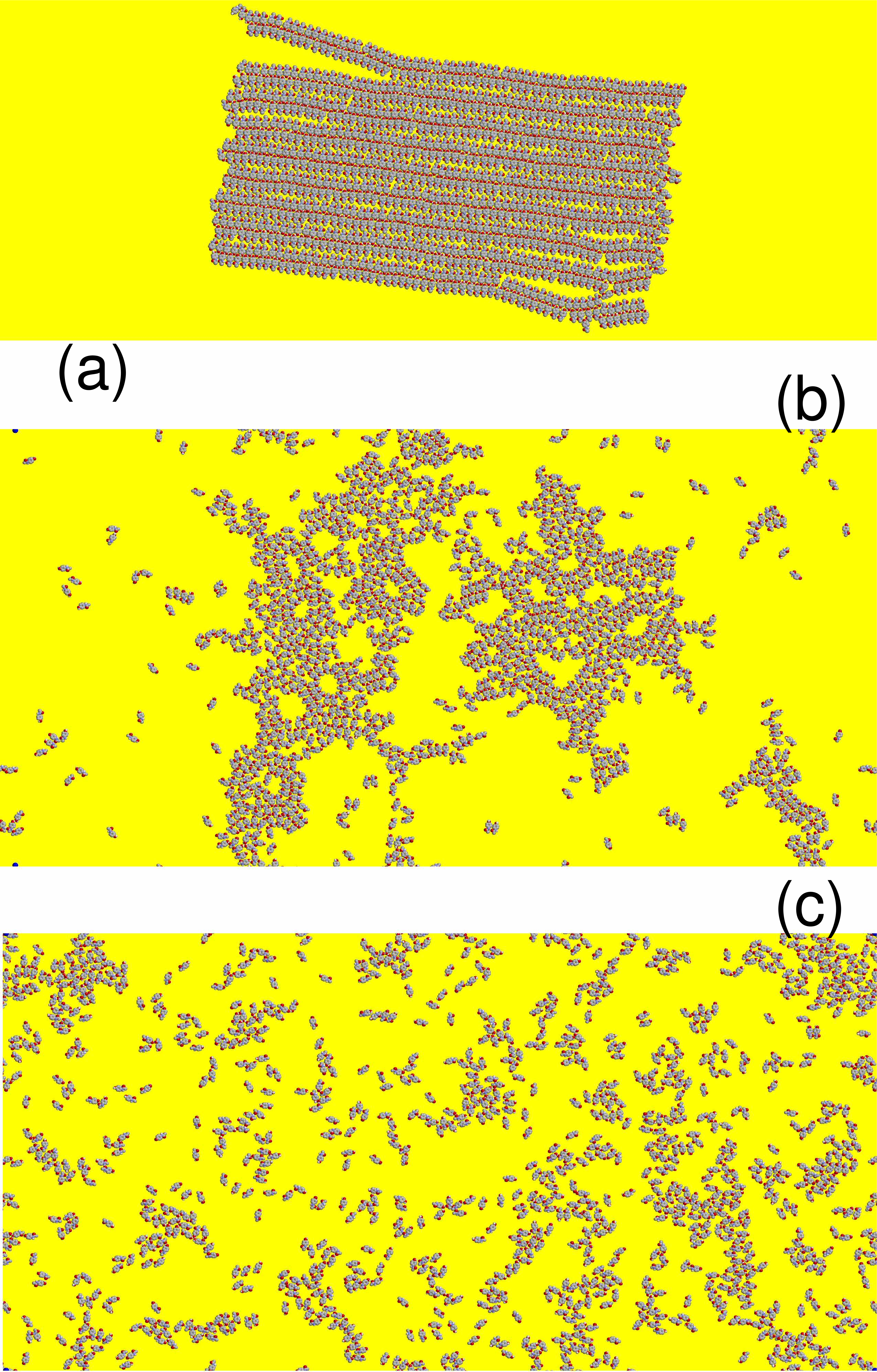}
\end{center}
\caption{\label{fig07}\protect
Structure of the 2D crystallite of $N=1080$ PCM molecules at 
(a) $T=320$~K (onset of melting, $p_{\mathrm{hh}}=0.92$),
(b) 430~K, and 
(c) 490~K (end of melting, $p_{\mathrm{hh}}=0.05$).
}
\end{figure}

During the time $t=2$~ns, the molecular system comes into equilibrium with the thermostat.
Further integration of the equations of motion (\ref{f16}) allows one to find the time-averaged values of the system energy $\overline{\!E}(T)$, the number of hydrogen bonds $\overline{\!N}_{\mathrm{hh}}(T)$, $\overline{\!N}_{\mathrm ah}(T)$, $\overline{\!N}_{\mathrm{aa}}(T)$, and the number of molecules lying on the substrate  $\overline{\!N}_{\mathrm s}(T)$.
For convenience of analyzing the dynamics results, we assume that the hydroxyl and amide groups of two molecules form hydrogen bonds type hg-hg, ag-hg, and ag-ag if the corresponding interaction energies satisfy $E_{\mathrm{hh}}<E_0$, $E_{\mathrm{ah}}<E_0$, and $E_{\mathrm{aa}}<E_0$, respectively, where $E_0=-0.11$~eV.
A molecule is considered to lie on the substrate if the distance from its center of mass to the substrate surface ($z=0$) does not exceed 9~\AA.
For the stationary state of the homochiral 2D crystal of $N=2N_xN_y$ molecules, the numbers of hydrogen bonds are $N_{\mathrm{hh}}=2N$, $N_{\mathrm{ah}}=0$, and $N_{\mathrm{aa}}=2N$; for the racemic one,
$N_{\mathrm{hh}}=0$, $N_{\mathrm{ah}}=4N$, and $N_{\mathrm{aa}}=0$, and $N_{\mathrm s}=N$ (see Fig.~\ref{fig04}).

The state of the molecular system at temperature $T$ can be characterized by its dimensionless heat capacity
\begin{equation}
c(T)=\frac{1}{3NN_0k_B}\frac{d\overline{\!E}}{dT},
\label{f17}
\end{equation}
the normalized numbers of hydrogen bonds $p_{\mathrm{hh}}(T)=\overline{\!N}_{\mathrm{hh}}/2N$,
$p_{\mathrm{ah}}(T)=\overline{\!N}_{\mathrm{ah}}/2N$, and $p_{\mathrm{aa}}(T)=\overline{\!N}_{\mathrm{aa}}/2N$,
and the fraction of molecules lying on the substrate $p_{\mathrm s}(T)=\overline{\!N}_{\mathrm s}/N$.
For an ideal 2D crystal at low temperature, the heat capacity is $c=1$, the normalized numbers of hydrogen bonds are $p_{\mathrm{hh}}=p_{\mathrm{aa}}=1$, and $p_{\mathrm{ah}}=0$ for the homochiral layer, and
$p_{\mathrm{hh}}=p_{\mathrm{aa}}=0$, $p_{\mathrm{ah}}=2$ for the racemic layer.
The fraction of molecules lying on the substrate is $p_{\mathrm s} =1$ (all molecules lie on the substrate at a distance of 3.3~\AA~ from its surface).
\begin{figure}[tb]
\begin{center}
\includegraphics[angle=0, width=1.0\linewidth]{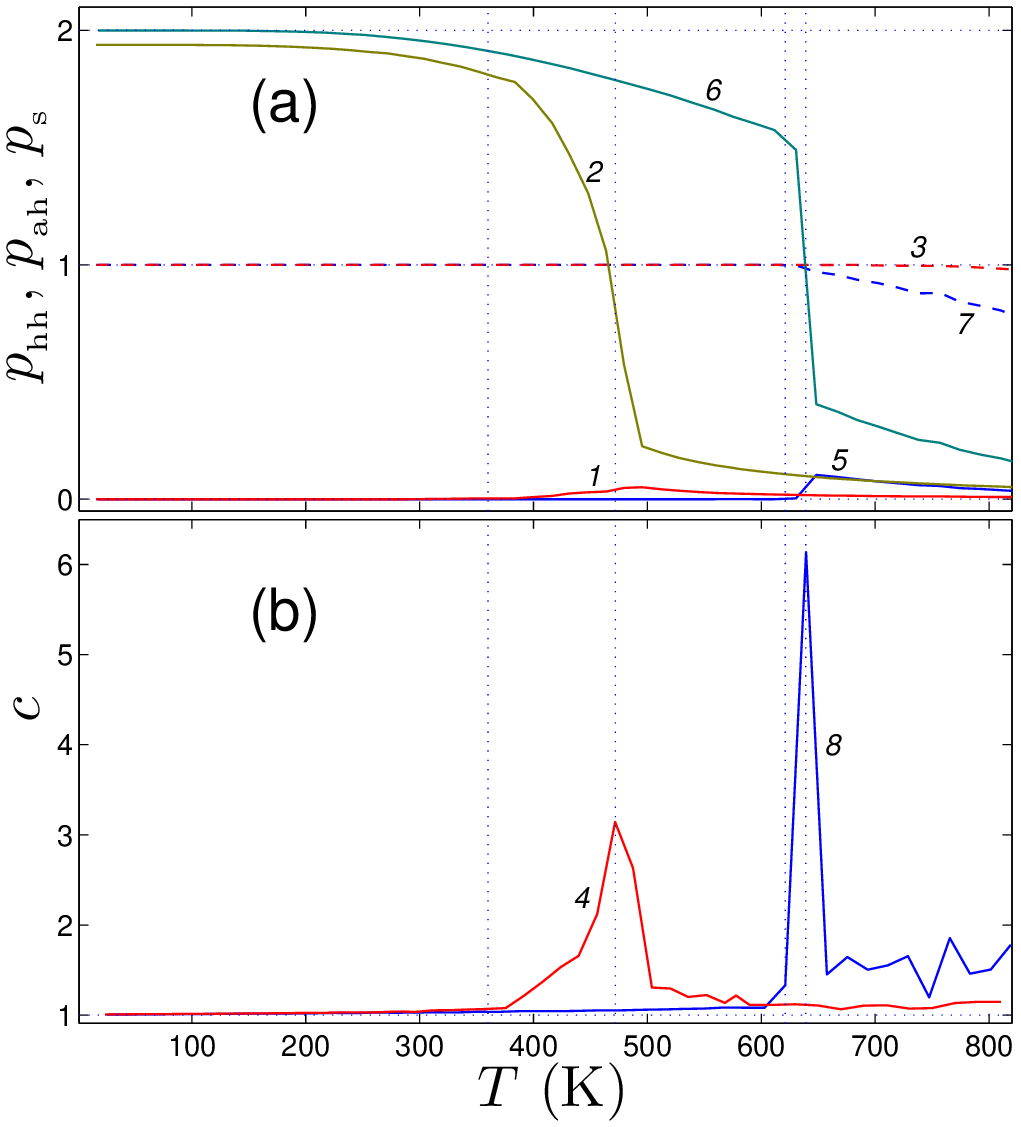}
\end{center}
\caption{\label{fig08}\protect
Temperature dependencies of (a) the normalized numbers of hydrogen bonds $p_{\mathrm{hh}}$ and $p_{\mathrm{ah}}$ (solid lines: 1, 5 for $p_{\mathrm{hh}}$  and 2, 6 for $p_{\mathrm{ah}}$) and the fraction $p_{\mathrm s}$ of molecules lying on the substrate (dashed lines: 3 and 7), and (b) the dimensionless heat capacity $c$ (curves 4 and 8), for the racemic structure of $N=1080$ PCM molecules.
Curves 1, 2, 3, and 4 correspond to a rectangular crystallite with free edges, covering 25~\% of the substrate, while curves 5, 6, 7, and 8 correspond to a two-dimensional crystal covering the entire substrate.
The vertical dotted lines mark the temperatures 360, 470, 620, and 640~K.
}
\end{figure}
\begin{figure}[tb]
\begin{center}
\includegraphics[angle=0, width=1.0\linewidth]{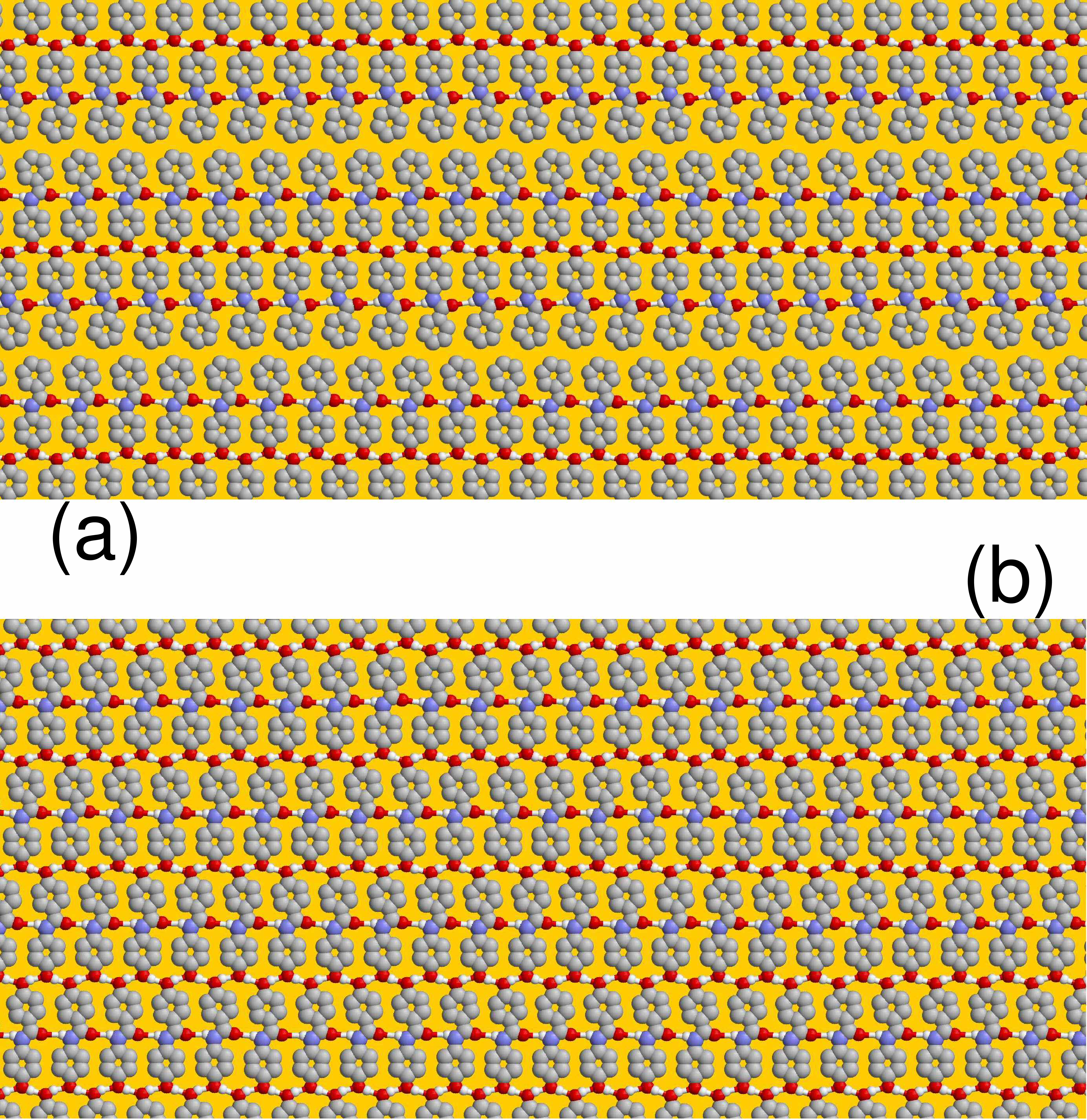}
\end{center}
\caption{\label{fig09}\protect
Crystal structures of the layers of (a) 4-hydroxybenzanilide (HBZL) and (b) 4,4$^\prime$-dihydroxybenzanilide (DHBZL) molecules on the flat surface of an h-BN crystal.
The color coding is the same as in Fig.~\ref{fig04}.
}
\end{figure}

The melting of 2D crystals differs substantially from that of 3D crystals.
Whereas for a 3D crystal melting is a first-order phase transition occurring at fixed temperature $T_0$, for a two-dimensional system it occurs continuously in the temperature interval $[T_1,T_2]$ \cite{Mak2006,Ryzhov2017,Tsiok2020,Toledano2021,Zhang2021}.
The dynamics simulations show that the melting of two-dimensional crystals of flat molecules on a flat substrate is also continuous.

The dependences of $c$, $p_{\mathrm{hh}}$, $p_{\mathrm{ah}}$, and $p_{\mathrm s}$ on temperature $T$ for a 2D crystallite and a crystal of homochiral PCM molecules are shown in Fig. \ref{fig06}.
Let us consider the melting of a two-dimensional crystallite covering 25\% of the substrate surface.
As seen in the figure, at low temperatures ($T<200$~K) the dimensionless heat capacity is $c=1$, the normalized numbers of hydrogen bonds are $p_{\mathrm{hh}}=1$ and $p_{\mathrm{ah}}=0$, and $p_{\mathrm s}=1$.
At these temperatures all hydrogen bonds are preserved, and all molecules are located on the substrate.
As the temperature increases, the number of hh hydrogen bonds gradually decreases, accompanied by a slight increase in the heat capacity.
Upon reaching $T_{1,{\mathrm{PCM}}}=315$~K, a noticeable increase in the heat capacity of the crystallite begins, accompanied by a rapid decrease in the number of hydrogen bonds.
This corresponds to the onset of melting of the crystallite, at which the edge chains begin to peel off from the crystallite (see Fig.~\ref{fig07}(a)).
A further increase in temperature leads to the rupture of long molecular chains and the formation of a porous structure of short chains with mixed hydrogen bonds (Fig.~\ref{fig07}~(b)).
Here the heat capacity of the molecular system increases rapidly, and the number of hydrogen bonds
between hydroxyl groups decreases (a decrease in $p_{\mathrm{hh}}$).
This is accompanied by the appearance and growth of mixed hydrogen bonds between hydroxyl and amide groups (an increase in $p_{\mathrm{ah}}$).
The maximum value of the heat capacity is reached at $T_{2,{\mathrm{PCM}}}=430$~K, after which it decreases rapidly and monotonically.
At $T>T_{2,{\mathrm{PCM}}}$, the short chains of mixed hydrogen bonds begin to break down.
The total number of bonds decreases monotonically, and the molecular system decomposes into individual molecules (see Fig. \ref{fig07}(c)). 
Note that at $T<720$~K all molecules remain on the flat substrate (the fraction $p_{\mathrm s}=1$).
Thus, the melting of the 2D crystallite of PCM molecules occurs continuously in the temperature interval
$[T_{1,{\mathrm{PCM}}},T_{2,{\mathrm{PCM}}}]$ and leads to a transition of the molecular system into a two-dimensional gas state.
\begin{figure}[tb]
\begin{center}
\includegraphics[angle=0, width=1.0\linewidth]{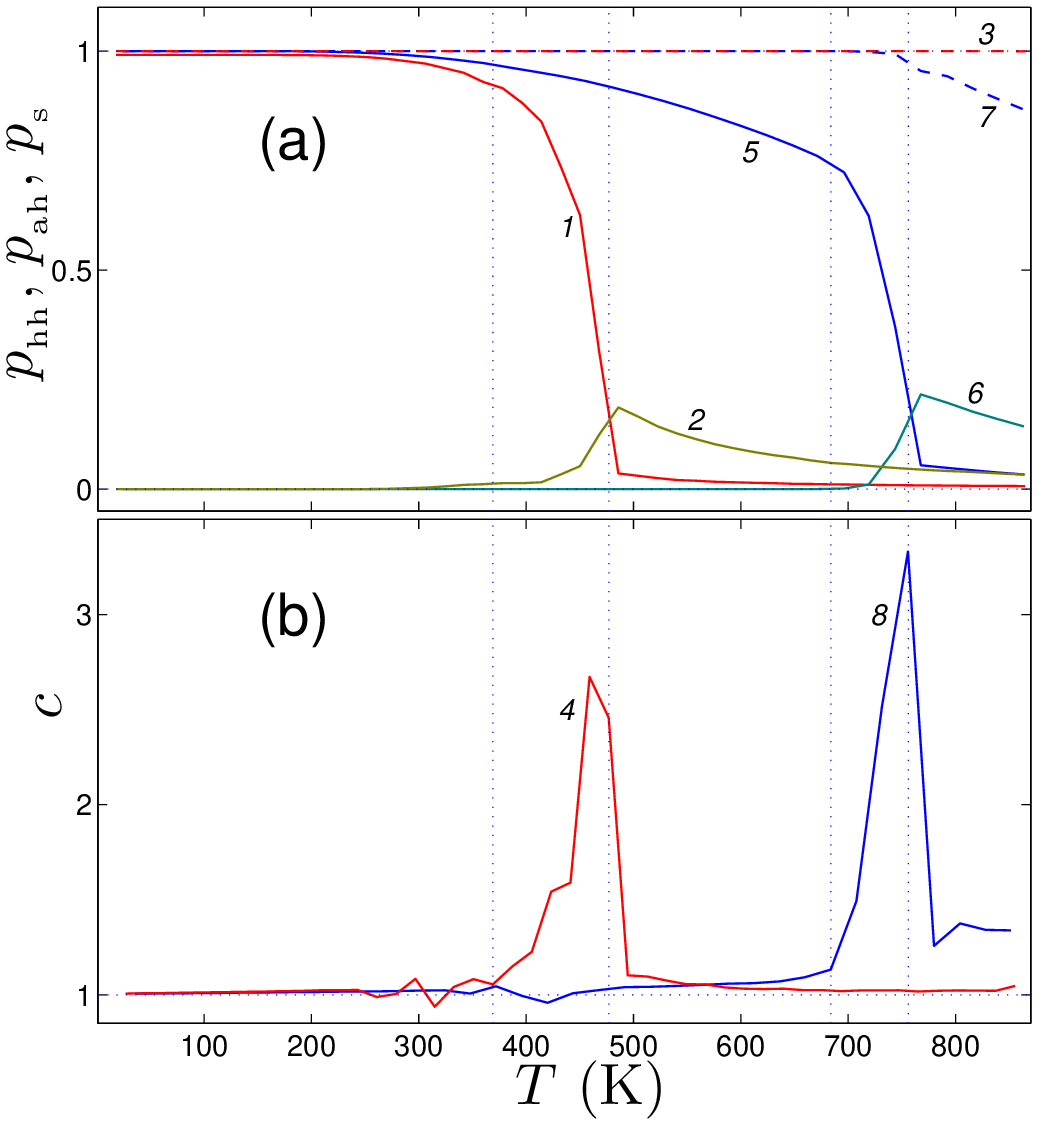}
\end{center}
\caption{\label{fig10}\protect
Temperature dependencies of (a) the normalized numbers of hydrogen bonds $p_{\mathrm{hh}}$ and $p_{\mathrm{ah}}$ (solid lines: 1, 5 for $p_{\mathrm{hh}}$  and 2, 6 for $p_{\mathrm{ah}}$) and the fraction $p_{\mathrm s}$ of molecules lying on the substrate (dashed lines: 3 and 7), and (b) the dimensionless heat capacity $c$ (curves 4 and 8), for the 2D structures of $N=1044$ HBZL molecules.
Curves 1, 2, 3, and 4 correspond to a rectangular crystallite with free edges, covering 25~\% of the substrate, while curves 5, 6, 7, and 8 correspond to a two-dimensional crystal covering the entire substrate.
The vertical dotted lines mark the temperatures 370, 480, 680, and 760~K.
}
\end{figure}
\begin{figure}[tb]
\begin{center}
\includegraphics[angle=0, width=0.97\linewidth]{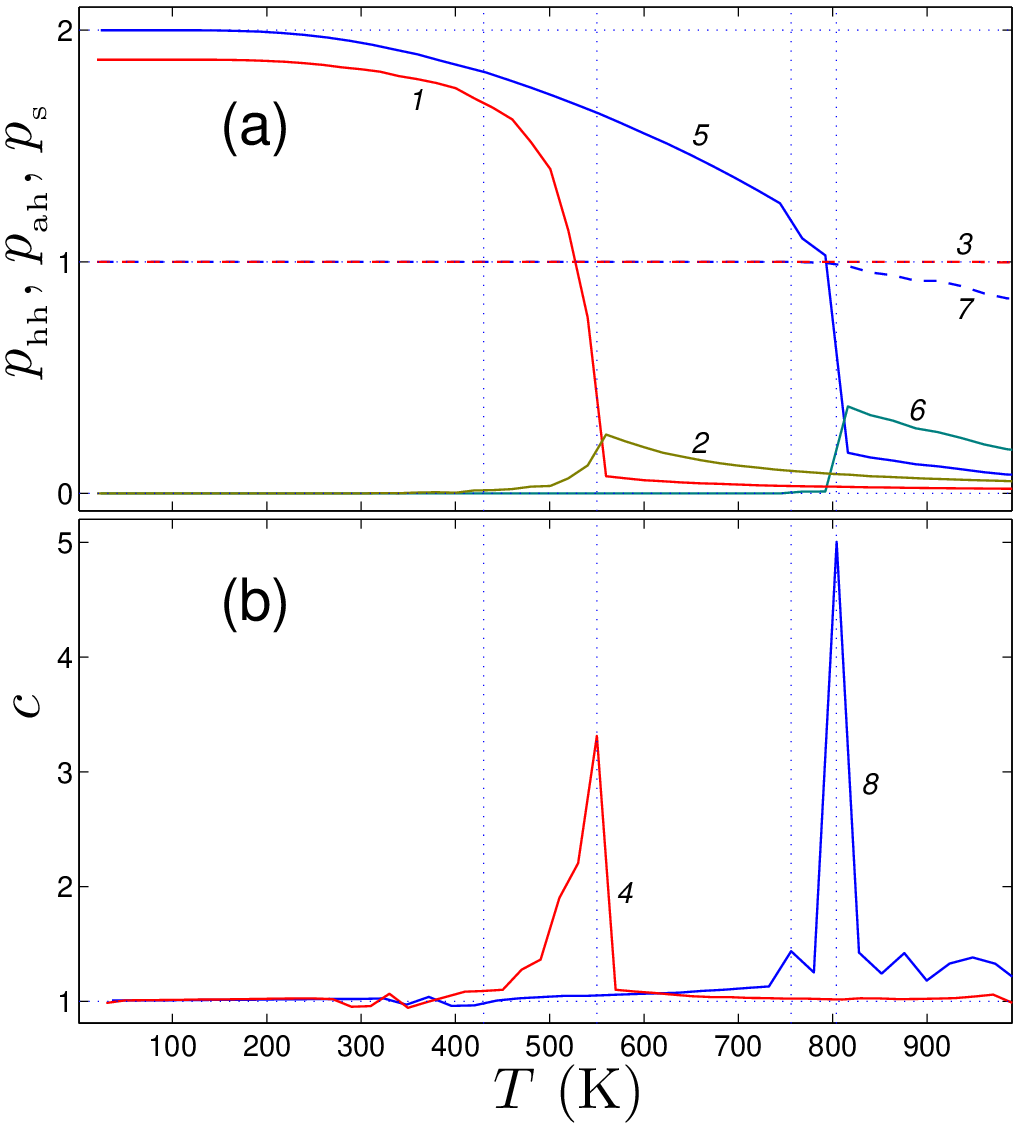}
\end{center}
\caption{\label{fig11}\protect
Temperature dependencies of (a) the normalized numbers of hydrogen bonds $p_{\mathrm{hh}}$ and $p_{\mathrm{ah}}$ (solid lines: 1, 5 for $p_{\mathrm{hh}}$  and 2, 6 for $p_{\mathrm{ah}}$) and the fraction $p_{\mathrm s}$ of molecules lying on the substrate (dashed lines: 3 and 7), and (b) the dimensionless heat capacity $c$ (curves 4 and 8), for the 2D structures of $N=1044$ DHBZL molecules.
Curves 1, 2, 3, and 4 correspond to a rectangular crystallite with free edges, covering 25~\% of the substrate, while curves 5, 6, 7, and 8 correspond to a two-dimensional crystal covering the entire substrate.
The vertical dotted lines mark the temperatures 430, 550, 756, and 804~K.
}
\end{figure}

For a 2D crystal covering the entire substrate, the heat capacity increases sharply and the number of hydrogen bonds decreases at $T\ge T_{3,{\mathrm{PCM}}}=550$~K.
Here, disordered regions with broken hydrogen bonds appear in the crystal (see Fig.~\ref{fig05}(b)).
Further melting of the crystal occurs through the growth of these regions, which leads to a sharp decrease in the number of hg-hg hydrogen bonds.
At $T_{4,{\mathrm{PCM}}}=570$~K, the melting of the 2D crystal ends, the heat capacity of the molecular system reaches its maximum value, and the regions with hydrogen-bonded chains completely disappear (see Fig.~\ref{fig05}(c)). 
Here most of the hydrogen bonds break, and part of the molecules detach from the substrate (a monotonic decrease in the fraction $p_{\mathrm s}$ begins).
The melting of the 2D crystal of PCM molecules occurs continuously in a narrower temperature interval
$[T_{3,{\mathrm{PCM}}},T_{4,{\mathrm{PCM}}}]$.
The melting is accompanied by the displacement of some molecules to a second layer, which subsequently leads to their detachment from the substrate.

Thus, the 2D crystallite of homochiral PCM molecules remains stable with respect to thermal vibrations at $T<315$~K, and the 2D crystal remains stable at $T<550$~K.

The two-dimensional racemic structure of PCM molecules (see Fig.~\ref{fig04}(b)) is more stable with respect to thermal vibrations than the homochiral one. 
For this structure, the continuous melting of a rectangular 2D crystallite of $N=1080$ molecules occurs in the temperature interval $[360,470]$~K, and that of the crystal in the interval $[620,640]$~K (see Fig.~\ref{fig08}).
Thus, the crystallite remains stable at $T<T_{1,{\mathrm{PCM}}}^*=360$~K, and the crystal at $T<T_{3,{\mathrm{PCM}}}^*=620$~K
(the asterisk denotes the racemic structure).
Note that the mixed hydrogen-bonded chains (\ref{f15}) typical of racemic two-dimensional structures of PCM molecules cannot be used for proton transport.
Proton transport requires hydrogen-bonded chains (\ref{f1}) consisting only of hydroxyl groups.
\begin{figure}[tb]
\begin{center}
\includegraphics[angle=0, width=1.0\linewidth]{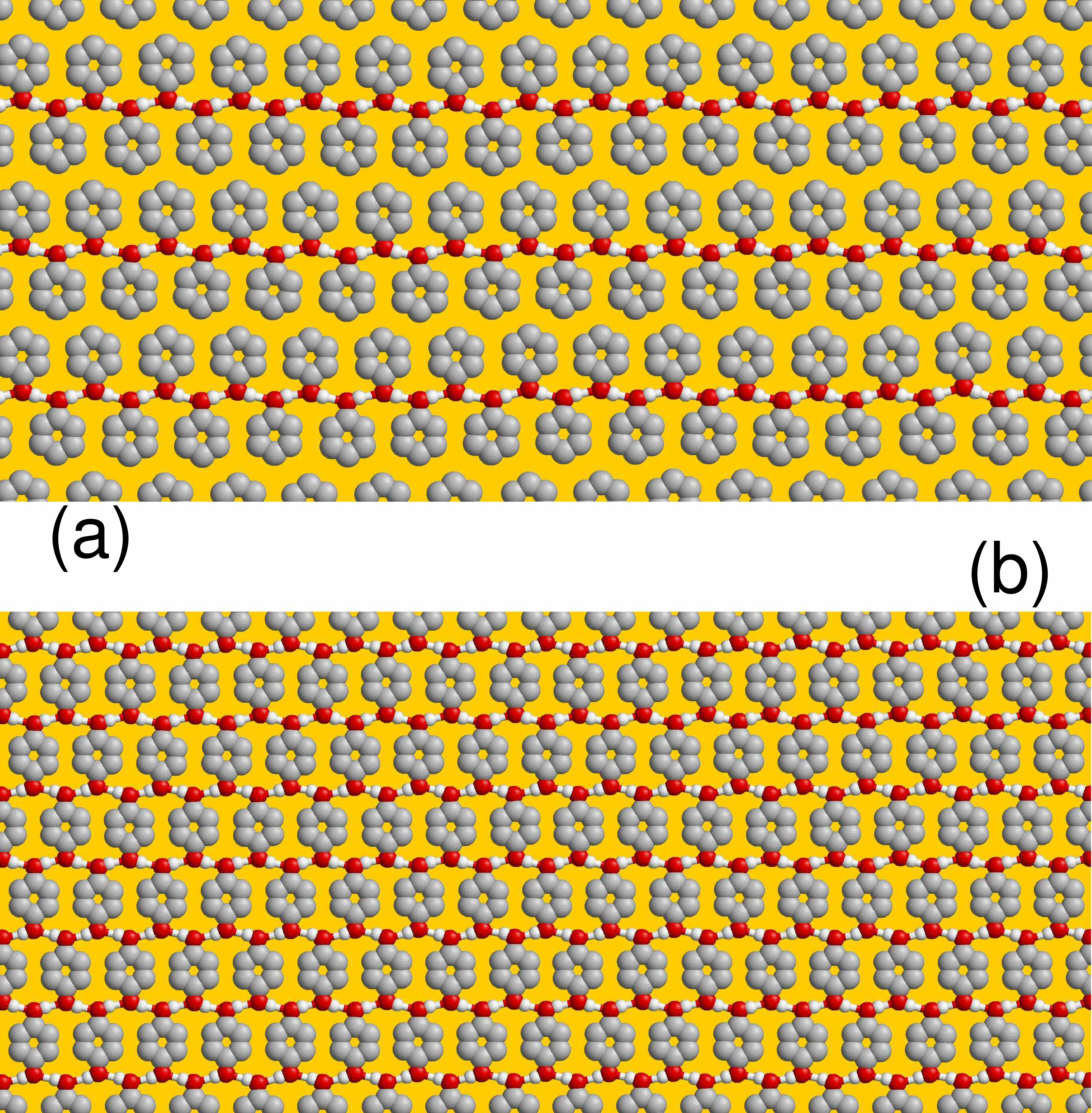}
\end{center}
\caption{\label{fig12}\protect
Crystal structure of the layers of (a) phenol and (b) hydroquinone molecules on the flat surface of an h-BN crystal.
The color coding is the same as in Fig.~\ref{fig04}.
}
\end{figure}

\section{Two-Dimensional Crystals of Hydroxybenzanilide Molecules}

The stability of two-dimensional molecular structures with hydrogen-bonded chains can be increased by enhancing the molecule-substrate interaction. 
To this end, the methyl group --CH$_3$ in the paracetamol molecule (PCM) can be replaced by a phenyl group --C$_6$H$_5$, yielding the molecule 4-hydroxybenzanilide C$_6$H$_5$C(O)NHC$_6$H$_4$OH (HBZL), which consists of $N_0=18$ united atoms (see Fig.~\ref{fig03}(e)). 
The stability of the two-dimensional structure can be further increased by increasing the number of possible hydrogen bonds. 
To this end, we can replace the methyl group in the PCM molecule with a hydroxyphenyl group --C$_6$H$_4$OH, obtaining the molecule 4,4$^\prime$-dihydroxybenzanilide C$_6$H$_4$OHC(O)NHC$_6$H$_4$OH (DHBZL), which consists of  $N_0=20$ united atoms (see Fig.~\ref{fig03}(f)). 
This molecule contains one amide group and two hydroxyl groups, so it can form six hydrogen bonds, and the two benzene rings ensure its strong interaction with the substrate.

Numerical minimization of the energy, Eq.~(\ref{f13}), showed that HBZL molecules on the flat h-BN substrate can form a flat periodic structure consisting of parallel molecular chains that form three hydrogen-bonded chains: two amide-group chains (\ref{f14}) and one hydroxyl-group chain (\ref{f1}) located between them (see Fig.~\ref{fig09}(a)).
The distance between neighboring oxygen atoms in the chain is $|\mathrm{OO}|=2.91$~\AA, the angle is $\angle\mathrm{OOO}=149^\circ$, and the hydrogen bond energy is $E_{\mathrm{hh},0}=-0.23$~eV.
The hydrogen-bonded chains of amide groups are linear, with a step $a=5.62$~\AA, and the bond energy is $E_{\mathrm{aa},0}=-0.22$~eV.
The unit cell of the structure consists of two molecules, and the lattice periods are $a_x=5.62$ and $a_y=24.54$~\AA.

To simulate the dynamics of such a structure, let us consider a 2D crystal of $N=2N_xN_y$ HBZL molecules, where the factor 2 accounts for the two molecules per unit cell.
The crystal  consists of $N_y=9$ parallel chains, each containing  $N_x=58$ units.
When periodic boundary conditions with periods $L_x=N_xa_x$, $L_y=N_ya_y$ are used, this structure models an infinite crystal covering the entire substrate.
When a periodic cell of size $L_x=2N_xa_x$, $L_y=2N_ya_y$ is used, a two-dimensional crystallite with free edges is modeled, covering only 25\% of the substrate surface.

Temperature dependences of the normalized numbers of hydrogen bonds $p_{\mathrm{hh}}$ and $p_{\mathrm{ah}}$, the fraction $p_{\mathrm s}$ of molecules lying on the substrate, and the dimensionless heat capacity $c$ for the 2D crystallite and crystal of HBZL molecules are shown in Fig.~\ref{fig10}.
As seen in the figure, the melting of a rectangular crystallite of $N=1044$ molecules occurs continuously in the temperature interval $[370,480]$~K, and that of the crystal in the interval $[680,760]$~K.
Thus, the crystallite of HBZL molecules remains stable at $T<T_{1,{\mathrm{HBZL}}}=370$~K, and the 2D crystal at $T<T_{3,{\mathrm{HBZL}}}=680$~K.
\begin{figure}[tb]
\begin{center}
\includegraphics[angle=0, width=1.0\linewidth]{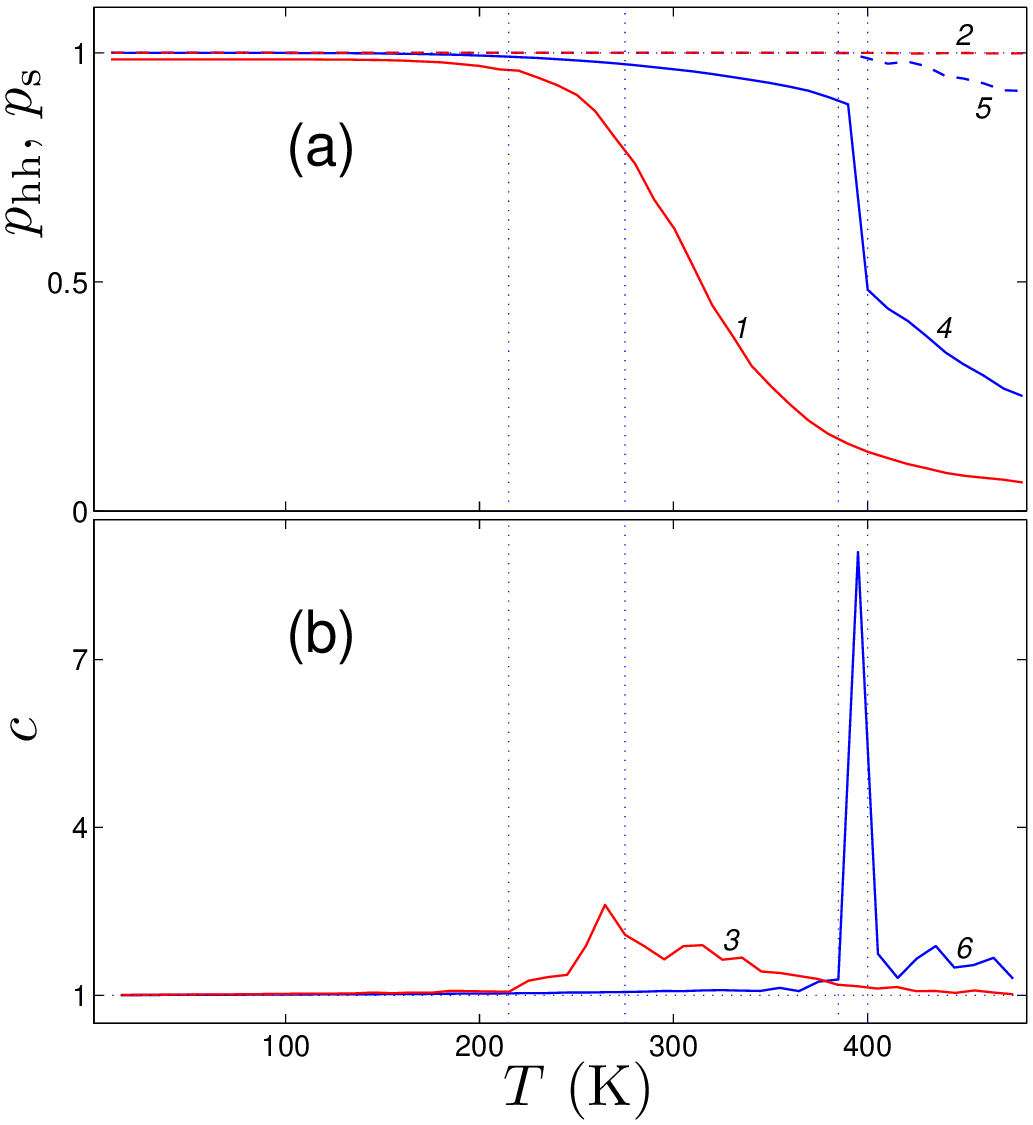}
\end{center}
\caption{\label{fig13}\protect
Temperature dependences of (a) the normalized number of hydrogen bonds $p_{\mathrm{hh}}$ (solid lines: 1 and 4) and the fraction $p_{\mathrm s}$ of molecules lying on the substrate (dashed lines: 2 and 5), and (b) the dimensionless heat capacity $c$ (curves 3 and 6), for the 2D structures of $N=1008$ PhOH molecules.
Curves 1, 2, and 3 correspond to a rectangular crystallite with free edges, covering 25~\% of the substrate, while curves 4, 5, and 6 correspond to a two-dimensional crystal covering the entire substrate.
The vertical dotted lines mark the temperatures 215, 275, 385, and 400~K.
}
\end{figure}
\begin{figure}[tb]
\begin{center}
\includegraphics[angle=0, width=1.0\linewidth]{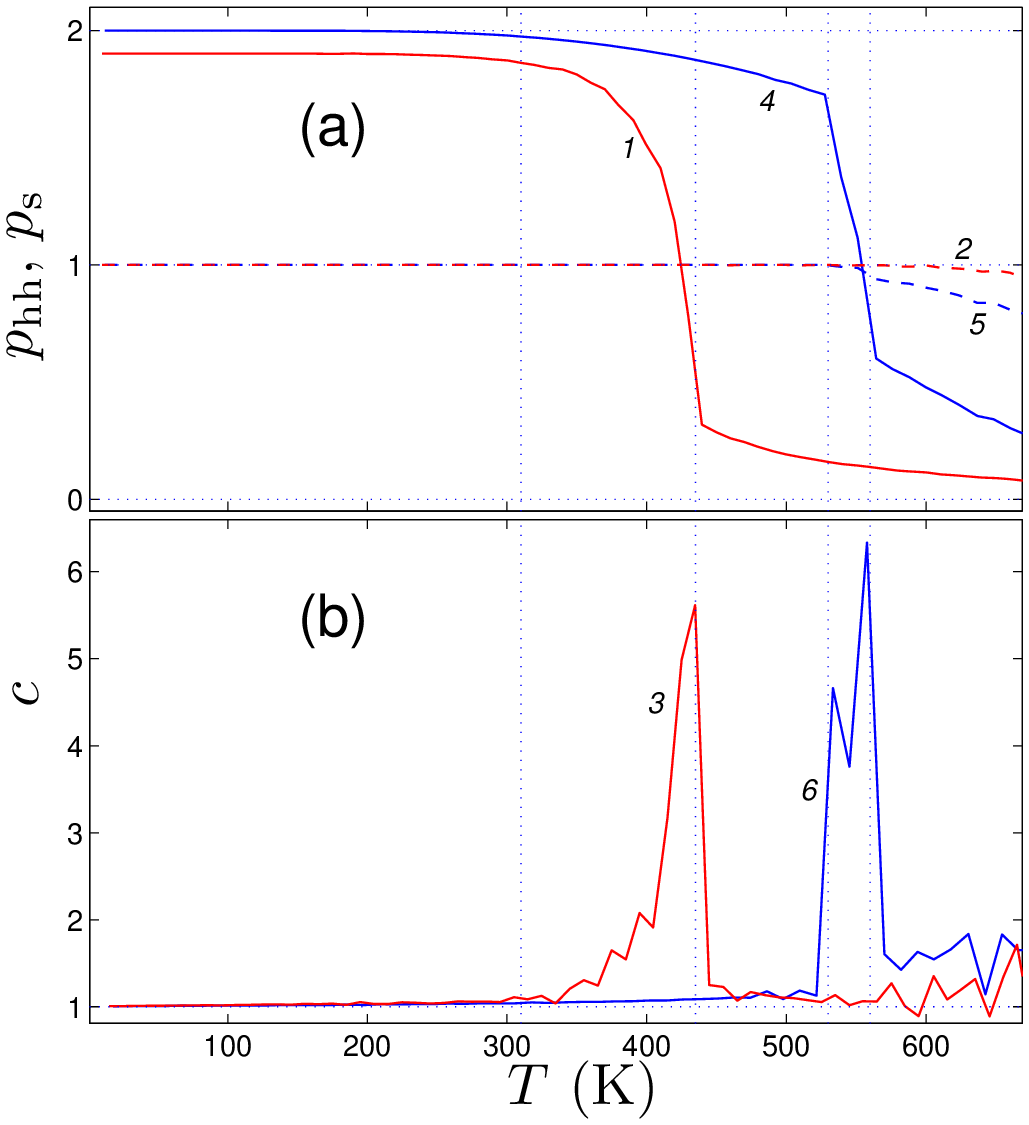}
\end{center}
\caption{\label{fig14}\protect
Temperature dependences of (a) the normalized number of hydrogen bonds $p_{\mathrm{hh}}$ (solid lines: 1 and 4) and the fraction $p_{\mathrm s}$ of molecules lying on the substrate (dashed lines: 2 and 5), and (b) the dimensionless heat capacity $c$ (curves 3 and 6), for the 2D structures of $N=1008$ HQ molecules.
Curves 1, 2, and 3 correspond to a rectangular crystallite with free edges, covering 25~\% of the substrate, while curves 4, 5, and 6 correspond to a two-dimensional crystal covering the entire substrate.
The vertical dotted lines mark the temperatures 310, 435, 530, and 560~K.
}
\end{figure}

Numerical minimization of the energy, Eq.~(\ref{f13}), showed that DHBZL molecules on the flat h-BN substrate form a periodic monolayer structure consisting of parallel hydrogen-bonded chains (see Fig.~\ref{fig09}(b)).
Here each molecule is linked by hydrogen bonds to all neighboring molecules. 
The unit cell of the two-dimensional lattice consists of two molecules, and the lattice periods are $a_x=5.56$ and $a_y=25.68$~\AA.
In the two-dimensional crystal, alternating parallel chains of hydrogen bonds (\ref{f1}) and (\ref{f14}) are formed.
Here the parameters of the hydrogen-bonded chains are
$|\mathrm{OO}|=2.92$~\AA, $\angle\mathrm{OOO}=144^\circ$, $a=a_x$, $E_{{\mathrm{hh}},0}=-0.225$~eV, and $E_{\mathrm{aa},0}=-0.21$~eV.

To simulate the dynamics of such a structure, let us consider a 2D crystal of $N=2N_xN_y$ DHBZL molecules ($N_x=58$, $N_y=9$), where the factor 2 accounts for the two molecules per unit cell.
Temperature dependences of $p_{\mathrm{hh}}$, $p_{\mathrm{ah}}$, $p_{\mathrm s}$, and $c$ for the 2D crystallite and crystal of DHBZL molecules are shown in Fig.~\ref{fig11}.
In this case, each molecule forms six hydrogen bonds (three incoming, three outgoing), so at low temperatures ($T<200$~K) the normalized numbers of hydrogen bonds are $p_{\mathrm{hh}}=2$, $p_{\mathrm{ah}}=0$, and $p_{\mathrm{aa}}=1$.
As seen in the figure, the melting of the 2D crystallite of $N=1044$ molecules occurs continuously in the temperature interval $[430,550]$~K, and that of the crystal in the interval $[756,804]$~K.
Thus, the 2D crystallite of DHBZL molecules remains stable at $T<T_{1,{\mathrm{DHBZL}}}=430$~K, and the 2D crystal remains stable at $T<T_{3,{\mathrm{DHBZL}}}=756$~K.

\section{Two-Dimensional Crystals of Phenol and Hydroquinone Molecules}

Flat molecules that do not contain amide groups, such as phenol C$_6$H$_5$OH (PhOH) and hydroquinone (1,4-dihydroxybenzene) C$_6$H$_4$(OH)$_2$ (HQ) (see Fig.~\ref{fig03} (a) and (b)), can also form continuous hydrogen-bonded chains (\ref{f1}) on a flat substrate.

The phenol molecule consists of $N_0=8$ united atoms, has one hydroxyl group, and therefore can form two hydrogen bonds.
The presence of a benzene ring ensures strong binding of the molecule to the flat substrate.
Numerical minimization of the energy, Eq.~(\ref{f13}), showed that PhOH molecules on the flat h-BN substrate can form a flat periodic structure with parallel hydrogen-bonded chains (see Fig.~\ref{fig12}(a)).
The parameters of the hydrogen-bonded chains are $|\mathrm{OO}|=2.93$~\AA, $\angle\mathrm{OOO}=150^\circ$, $E_{{\mathrm{hh}},0}=-0.23$~eV.
The unit cell of the 2D structure consists of two molecules, and the lattice periods are $a_x=5.68$ and $a_y=11.41$~\AA.

To simulate the dynamics of such a structure, let us consider a 2D crystal of $N=2N_xN_y$ PhOH molecules, where the factor 2 accounts for the two molecules per unit cell.
The crystal consists of $N_y=14$ parallel chains, each containing $N_x=36$ units.
When periodic boundary conditions with periods $L_x=N_xa_x$ and $L_y=N_ya_y$ are used, this structure models an infinite crystal covering the entire substrate.
When a periodic cell of size $L_x=2N_xa_x$, $L_y=2N_ya_y$ is used, a two-dimensional crystallite with free edges is modeled, covering only 25\% of the substrate surface.

Temperature dependences of the normalized number of hydrogen bonds $p_{\mathrm{hh}}$, the fraction $p_{\mathrm s}$ of molecules lying on the substrate, and the dimensionless heat capacity $c$ for the 2D crystallite and crystal of phenol molecules are shown in Fig.~\ref{fig13}.
As seen in the figure, the melting of a rectangular crystallite of $N=1008$ molecules occurs continuously in the temperature interval $[215,275]$~K, and that of the 2D crystal occurs in the interval $[385,400]$~K.
Thus, two-dimensional structures of phenol molecules have low stability with respect to thermal vibrations:
the crystallite remains stable at $T<T_{1,{\mathrm{PhOH}}}=215$~K, and the 2D crystal remains stable at $T<T_{3,{\mathrm{PhOH}}}=385$~K.

The hydroquinone molecule consists of $N_0=10$ united atoms, has two hydroxyl groups, and therefore can form four hydrogen bonds.
The presence of a benzene ring ensures strong binding to the flat substrate.
Numerical minimization of the energy, Eq.~(\ref{f13}), showed that HQ molecules on the flat h-BN substrate can form a two-dimensional periodic structure with parallel hydrogen-bonded chains, in which each molecule forms hydrogen bonds with four neighbors (see Fig.~\ref{fig12}(b)).
The parameters of the hydrogen-bonded chains are $|\mathrm{OO}|=2.90$~\AA, $\angle\mathrm{OOO}=149^\circ$, $E_{{\mathrm{hh}},0}=-0.23$~eV.
The unit cell of the 2D structure consists of two molecules, and the lattice periods are $a_x=5.60$ and $a_y=12.58$~\AA.
\begin{figure}[tb]
\begin{center}
\includegraphics[angle=0, width=0.90\linewidth]{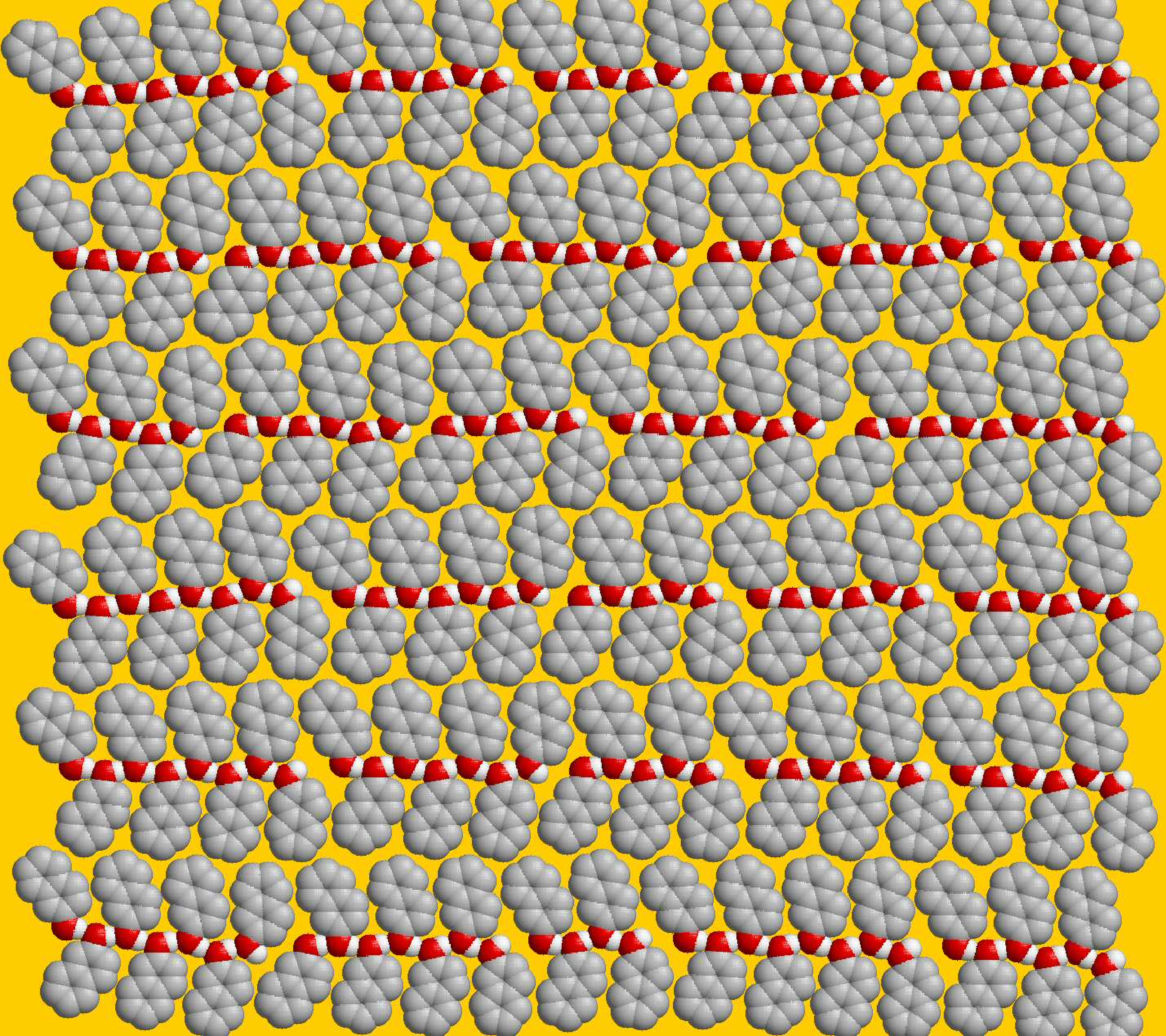}
\end{center}
\caption{\label{fig15}\protect
Crystal structure of the layer of $\beta$-naphthol molecules on the flat surface of an h-BN crystal.
The color coding is the same as in Fig.~\ref{fig04}.
}
\end{figure}
\begin{figure}[tb]
\begin{center}
\includegraphics[angle=0, width=1.0\linewidth]{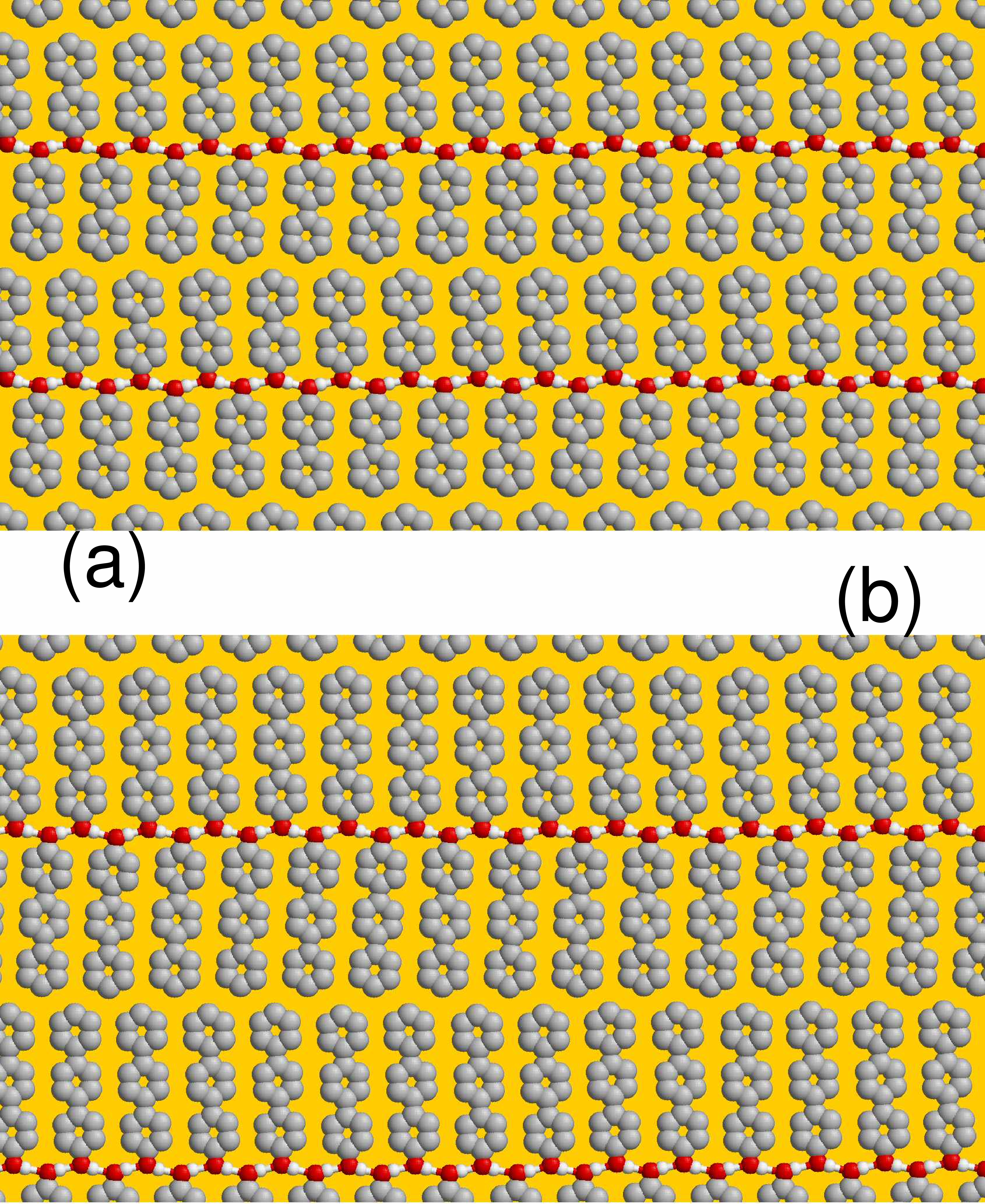}
\end{center}
\caption{\label{fig16}\protect
Crystal structure of the layer of (a) 4-phenylphenol (4PhPh) and (b) 4-(4-phenylphenyl)phenol (44PhPhPh) molecules located on the flat surface of an h-BN crystal.
}
\end{figure}

To simulate the dynamics, let us consider a 2D crystal of $N=2N_xN_y=1008$ HQ molecules ($N_x=36$, $N_y=14$), where the factor 2 accounts for the two molecules per unit cell.
The results of the dynamics simulations for 2D structures of hydroquinone molecules are presented in Fig.~\ref{fig14}.
As seen in the figure, the melting of the rectangular crystallite occurs continuously in the temperature interval $[310,435]$~K, and that of the crystal occurs in the interval $[530,560]$~K.
Thus, 2D structures of hydroquinone molecules have higher thermal stability than those of phenol: the crystallite remains stable at $T<T_{1,{\mathrm{HQ}}}=310$~K, and the crystal remains stable at $T<T_{3,{\mathrm{HQ}}}=530$~K.

The thermal stability of 2D structures can be further increased by enhancing the molecule-substrate interaction. 
To this end, additional benzene rings can be introduced into the molecule. 
However, if a new ring is fused to an existing one, forming a single flat aromatic system, the size of the molecule does not allow long continuous hydrogen-bonded chains to form on a flat substrate; steric effects hinder the formation of such chains.
For example, $\beta$-naphthol C$_{10}$H$_7$OH molecules on a flat substrate can form continuous linear
chains of at most 8 units (see Fig.~\ref{fig15}).

An increase in the transverse dimension of the molecule can be avoided if the benzene ring is added via a C--C valence bond, forming the molecules 4-phenylphenol C$_6$H$_5$--C$_6$H$_4$OH (4PhPh) and 4-(4-phenylphenyl)phenol C$_6$H$_5$--C$_6$H$_4$--C$_6$H$_4$OH (44PhPhPh) (see Fig.~\ref{fig03}(c) and (d)).
The 4PhPh molecule consists of $N_0=14$ united atoms, and 44PhPhPh of $N_0=20$
(the first molecule contains two benzene rings, the second three).
Numerical minimization of the energy, Eq.~(\ref{f13}), showed that the 4PhPh and 44PhPhPh molecules, like phenol, can form flat periodic structures with continuous parallel hydrogen-bonded chains on the flat h-BN substrate (see Fig.~\ref{fig16}).
The periods of the two-dimensional lattice are $a_x=5.70$ and $a_y=19.93$~\AA~ for 4PhPh molecules and
$a_x=5.70$ and $a_y=28.47$~\AA~ for 44PhPhPh.
The parameters of the hydrogen-bonded chain are $|\mathrm{OO}|=2.94$~\AA, $\angle\mathrm{OOO}=152^\circ$, $E_{\mathrm{hh},0}=-0.21$~eV
for the 2D lattice of 4PhPh and 44PhPhPh molecules.
\begin{figure}[tb]
\begin{center}
\includegraphics[angle=0, width=1.0\linewidth]{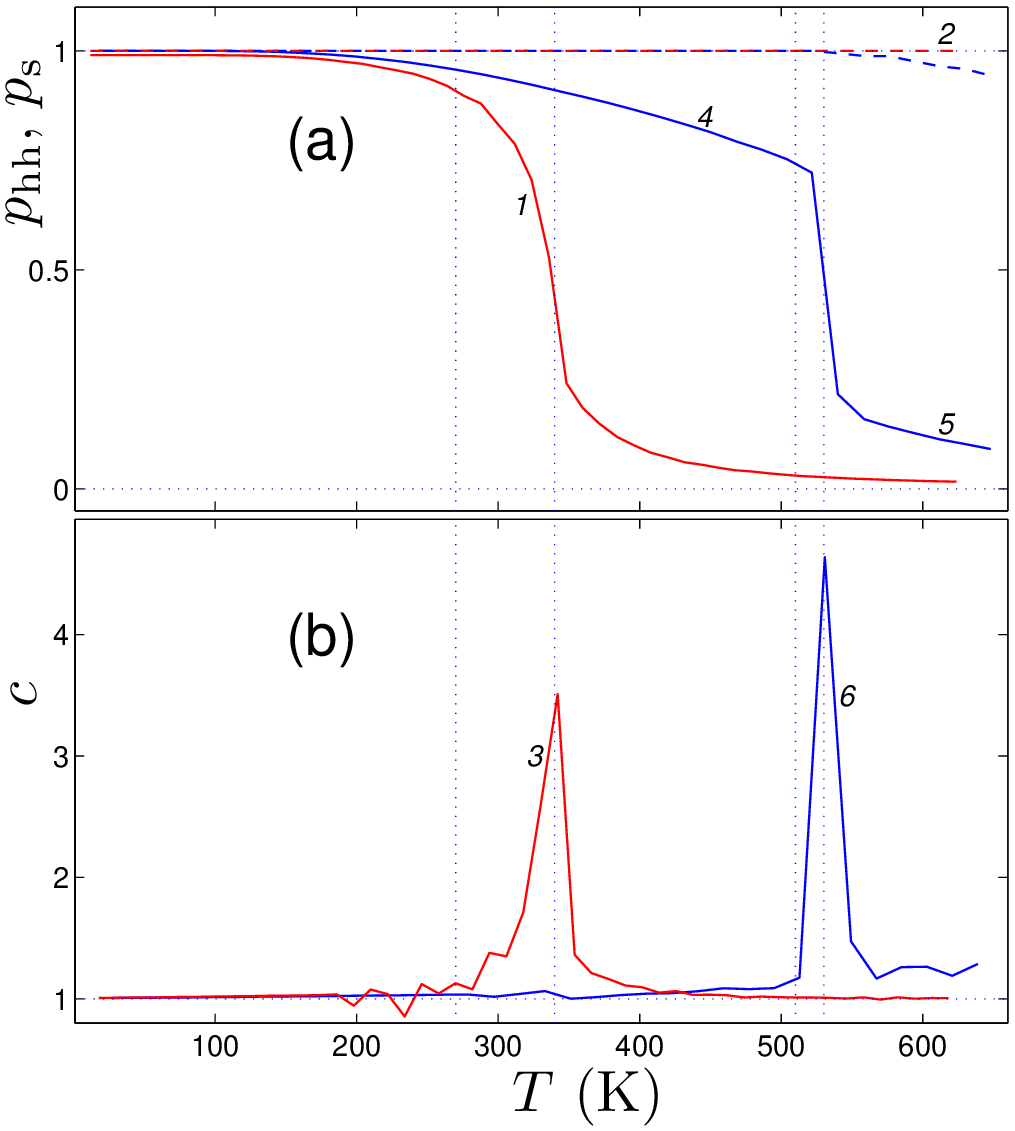}
\end{center}
\caption{\label{fig17}\protect
Temperature dependences of (a) the normalized number of hydrogen bonds $p_{\mathrm{hh}}$ (solid lines: 1 and 4) and the fraction $p_{\mathrm s}$ of molecules lying on the substrate (dashed lines: 2 and 5), and (b) the dimensionless heat capacity $c$ (curves 3 and 6), for the 2D structures of $N=1400$ 4PhPh molecules.
Curves 1, 2, and 3 correspond to a rectangular crystallite with free edges, covering 25\% of the substrate, while curves 4, 5, and 6 correspond to a two-dimensional crystal covering the entire substrate.
The vertical dotted lines mark the temperatures 270, 340, 510, and 530~K.
}
\end{figure}
\begin{figure}[tb]
\begin{center}
\includegraphics[angle=0, width=1.0\linewidth]{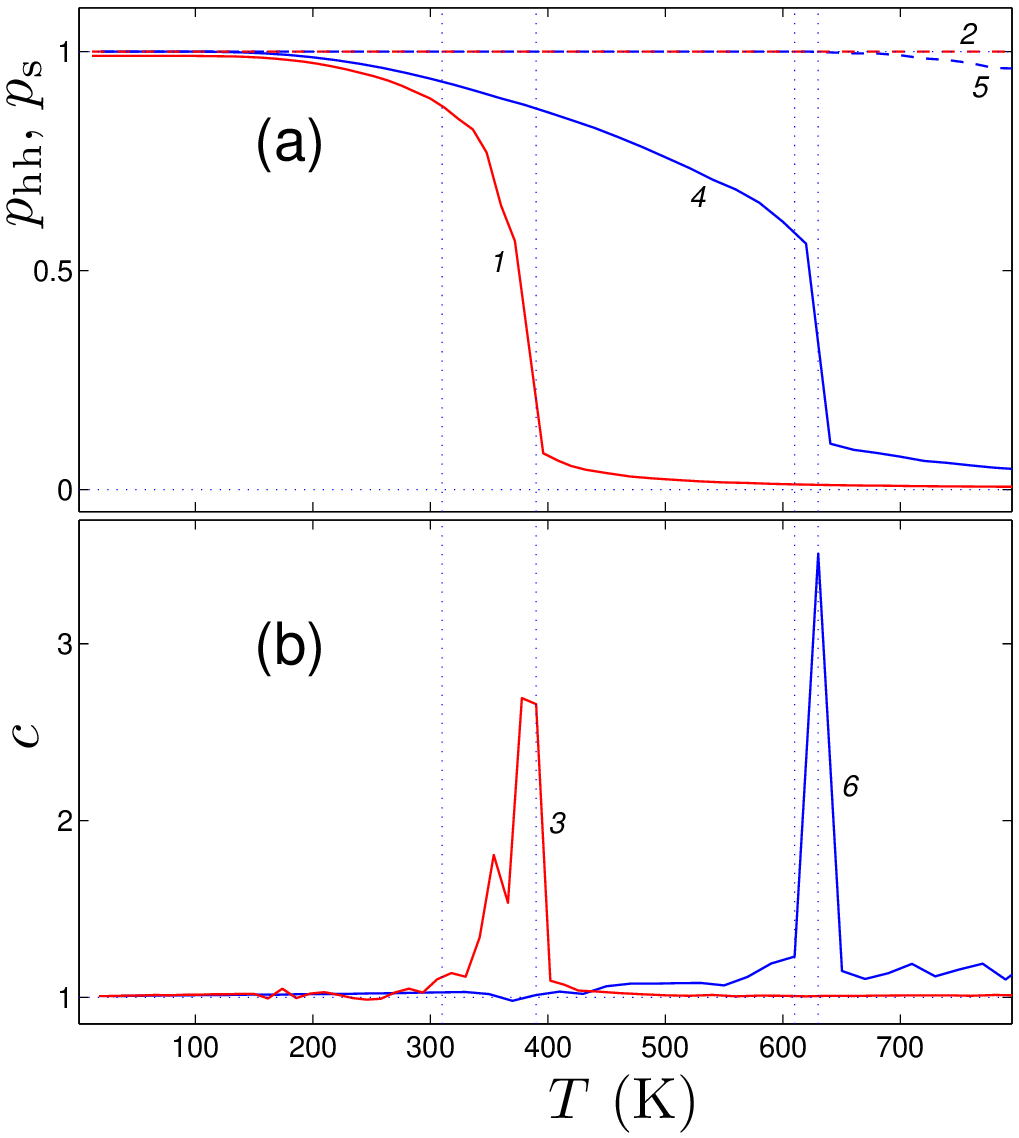}
\end{center}
\caption{\label{fig18}\protect
Temperature dependences of (a) the normalized number of hydrogen bonds $p_{\mathrm{hh}}$ (solid lines: 1 and 4) and the fraction $p_{\mathrm s}$ of molecules lying on the substrate (dashed lines: 2 and 5), and (b) the dimensionless heat capacity $c$ (curves 3 and 6), for the 2D structures of $N=1000$ 44PhPhPh molecules.
Curves 1, 2, and 3 correspond to a rectangular crystallite with free edges, covering 25\% of the substrate, while curves 4, 5, and 6 correspond to a two-dimensional crystal covering the entire substrate.
The vertical dotted lines mark the temperatures 310, 390, 610, and 630~K.
}
\end{figure}

To simulate the dynamics of such a 2D lattice of 4PhPh molecules, let us consider a crystal of $N=2N_xN_y=1400$ molecules ($N_x=50$, $N_y=14$), where the factor 2 accounts for the two molecules per unit cell.
The results of the dynamics simulations for 2D structures of 4PhPh molecules are presented in Fig.~\ref{fig17}.
As seen in the figure, the melting of the rectangular crystallite occurs continuously in the temperature interval [270,340]~K, and that of the crystal occurs in the interval [510,530]~K.
Thus, 2D structures of 4PhPh molecules have higher thermal stability than those of phenol: the crystallite remains stable at $T<T_{1,{\mathrm{4PhPh}}}=270$~K, and the crystal remains stable at $T<T_{3,{\mathrm{4PhPh}}}=510$~K.

To simulate the dynamics of 2D lattices of 44PhPhPh molecules, let us consider a crystal of $N=2N_xN_y=1000$ molecules ($N_x=50$, $N_y=10$), where the factor 2 accounts for the two molecules per unit cell. 
The results of the dynamics simulations for 2D structures of 44PhPhPh molecules are presented in Fig.~\ref{fig18}.
As seen in the figure, the melting of the rectangular crystallite occurs continuously in the temperature interval [310,390]~K, and that of the crystal occurs in the interval [610,630]~K.
Thus, 2D structures of 44PhPhPh molecules have even higher thermal stability than those of 4PhPh: the crystallite remains stable at $T<T_{1,{\mathrm{44PhPhPh}}}=310$~K,
and the crystal remains stable at $T<T_{3,{\mathrm{44PhPhPh}}}=610$~K.\\

\section{Conclusion}

Numerical simulations of the dynamics of monolayer structures of molecules deposited on a sheet of hexagonal boron nitride (h-BN) show that molecules containing benzene rings and hydroxyl groups can form stable two-dimensional crystals with linear hydrogen-bonded chains (\ref{f1}).
Such structures are formed by phenol, hydroquinone, 4-phenylphenol, 4-(4-phenylphenyl)phenol,
paracetamol, 4-hydroxybenzanilide, and 4,4$^\prime$-dihydroxybenzanilide.
The benzene rings in these molecules, on the one hand, ensure their strong interaction with the flat substrate, and on the other hand, do not hinder the formation of long hydrogen-bonded chains.
Such two-dimensional structures can exhibit high stability with respect to thermal vibrations: the onset melting temperatures of the 2D crystals of the listed molecules are
112, 257, 237, 337, 277, 407, and 483$^\circ$C, respectively.
Thus, in these structures the hydrogen-bonded chains can be preserved up to fairly high temperatures. Note that the melting temperature of phosphoric acid, most commonly used in proton-exchange membranes (PEMs), is $42^\circ$C.
Hydrogen-bonded chains are efficient pathways for proton transfer.
Therefore, the flat structures considered here can be used for the development of anhydrous PEMs with higher thermal stability than phosphoric-acid-based membranes.

To date, PEMs operating at temperatures up to 250$^\circ$C have been developed \cite{He2026,Stepanov2026}.
In these membranes, proton transport occurs along hydrogen-bonded chains formed by phosphoric acid molecules.
The simulations performed here allow us to conclude that structures composed of h-BN sheets and molecules of hydroquinone, paracetamol, 4-hydroxybenzanilide, and 4,4$^\prime$-dihydroxybenzanilide can be used
for the development of novel proton-exchange membranes capable of operating at elevated temperatures.\\

{\bf Acknowledgements}\\

Computational facilities were provided by the Joint Supercomputer Center (JSCC) of the National Research Center ``Kurchatov Institute''.
The research was funded by the Russian Science Foundation (RSF) (project No. 25-73-20038).

\end{document}